\documentclass[aps,pre,twocolumn,groupedaddress]{revtex4-2}
\usepackage{amsmath}
\usepackage[dvipdfmx]{graphicx}
\usepackage{color}
\usepackage{comment}
\begin{document}

% Use the \preprint command to place your local institutional report
% number in the upper righthand corner of the title page in preprint mode.
% Multiple \preprint commands are allowed.
% Use the 'preprintnumbers' class option to override journal defaults
% to display numbers if necessary
%\preprint{}

%Title of paper
\title{Compatibility of Carnot efficiency with finite power in an underdamped Brownian Carnot cycle in small temperature-difference regime}

% repeat the \author .. \affiliation  etc. as needed
% \email, \thanks, \homepage, \altaffiliation all apply to the current
% author. Explanatory text should go in the []'s, actual e-mail
% address or url should go in the {}'s for \email and \homepage.
% Please use the appropriate macro foreach each type of information

% \affiliation command applies to all authors since the last
% \affiliation command. The \affiliation command should follow the
% other information
% \affiliation can be followed by \email, \homepage, \thanks as well.
\author{Kosuke Miura, Yuki Izumida,$^{\dagger}$ and Koji Okuda}
%\email[]{miura@statphys.sci.hokudai.ac.jp}
%\homepage[]{Your web page}
%\thanks{}
%\altaffiliation{}
\affiliation{Department of Physics, Hokkaido University Sapporo 060-0810, Japan\\
  $^{\dagger}$Department of Complexity Science and Engineering, Graduate School of Frontier Sciences, The University of Tokyo, Kashiwa 277-8561, Japan}
%\footnote{miura@statphys.sci.hokudai.ac.jp}
%Collaboration name if desired (requires use of superscriptaddress
%option in \documentclass). \noaffiliation is required (may also be
%used with the \author command).
%\collaboration can be followed by \email, \homepage, \thanks as well.
%\collaboration{}
%\noaffiliation

\date{\today}

\begin{abstract}
  % insert abstract here
  We study the possibility of achieving the Carnot efficiency
  in a finite-power underdamped Brownian Carnot cycle.
  Recently, it was reported that the Carnot efficiency is achievable in a general class of finite-power Carnot cycles in the vanishing limit of the relaxation times.
  Thus, it may be interesting to clarify how the efficiency and power depend on the relaxation times by using a specific model.
  By evaluating the heat-leakage effect intrinsic in the underdamped dynamics with the instantaneous adiabatic processes, we demonstrate that the compatibility of the Carnot efficiency and finite power is achieved in the vanishing limit of the relaxation times in the small temperature-difference regime.  Furthermore, we show that this result is consistent with a trade-off relation between power and efficiency by explicitly deriving the relation of our cycle in terms of the relaxation times.
\end{abstract}

% insert suggested keywords - APS authors don't need to do this
%\keywords{}

%\maketitle must follow title, authors, abstract, and keywords
\maketitle

% body of paper here - Use proper section commands
% References should be done using the \cite, \ref, and \label commands
\section{Introduction}
Heat engines constitute one of the indispensable technologies in our modern society, and much effort has been conducted to improve their performance in various scientific or engineering fields \cite{Callen_2nd_book}.
Heat engines convert supplied heat into output work. Moreover, their ratio can be used as the efficiency to characterize the performance of heat engines.
The Carnot cycle is one of the most important models of heat engines, which operates between hot and cold heat baths
with constant temperatures $T_{h}$ and $T_{c}$ ($<T_{h}$). Moreover, the cycle is composed of two isothermal processes and two adiabatic processes.
Carnot demonstrated that the efficiency of any heat engine is limited by the upper bound called the Carnot efficiency~\cite{Carnot_book}:
\begin{equation}
  \label{Carnot efficiency}
  \eta_{C} \equiv 1-\frac{T_{c}}{T_{h}}.
\end{equation}
It is known that we can reach the Carnot efficiency by the reversible cycle,
where the heat engine always remains at equilibrium and
is typically operated quasistatically, which implies that
the engine spends an infinitely long time per cycle.
Moreover, power, defined as output work per unit time,
is another important quantity for evaluating the performance of heat engines.
When we operate the heat engines quasistatically, power vanishes.
Thus, several studies have been devoted to investigating the feasibility of finite-power heat engines with Carnot efficiency
\cite{Polettini_2017,PhysRevE.62.6021,PhysRevE.95.052128,
  PhysRevE.62.6021,Campisi2016,PhysRevLett.106.230602,
  PhysRevLett.110.070603,PhysRevB.87.165419,PhysRevB.94.121402,
  PhysRevLett.112.140601,PhysRevLett.114.146801,Sothmann_2014,PhysRevE.98.042112,PhysRevLett.124.110606,PhysRevE.96.062107}.

However, Shiraishi {\it et al}.
\cite{PhysRevLett.117.190601,PhysRevE.96.022138,N.Shiraishi2018}
recently proved a trade-off relation between power $P$ and efficiency $\eta$ in general heat engines described by the Markov process.
The trade-off relation is given by
\begin{equation}
  \label{general trade-off relation}
  P \leq A \eta (\eta_{C} - \eta),
\end{equation}
where $A$ is a positive constant depending on the heat engine details.
Based on this relation, the power should vanish as the efficiency approaches the Carnot efficiency.
Similar trade-off relations to
Eq.~(\ref{general trade-off relation}) have been obtained in
various heat engine models~\cite{PhysRevLett.120.190602,Koyuk_2018,Dechant_2018,PhysRevE.97.062101}.
In particular, Dechant and Sasa derived a specific expression
of $A$ for stochastic heat engines described by
the Langevin equation~\cite{PhysRevE.97.062101}.

Recently, Holubec and Ryabov reported that the Carnot efficiency could be obtained in a general class of finite-power Carnot cycles in the vanishing limit of the relaxation times~\cite{PhysRevLett.121.120601}.
Although this result seems to contradict the trade-off relation in Eq.~(\ref{general trade-off relation}),
they pointed out the possibility that $A$ in Eq.~(\ref{general trade-off relation}) diverges in the vanishing limit of the relaxation times,  and the Carnot efficiency and finite power are compatible without breaking the trade-off relation in Eq.~(\ref{general trade-off relation}).
Thus, it may be interesting to study how the efficiency and power depend on the relaxation times in more detail by using a specific model.

The Brownian Carnot cycle with instantaneous adiabatic processes and a time-dependent harmonic potential is a simple model, which is easy to analyze and is frequently used to study the efficiency and power~\cite{PhysRevLett.121.120601,Dechant_2017,Schmiedl_2007,PhysRevE.97.022131}.
However, it is pointed out that the instantaneous adiabatic process in the overdamped Brownian Carnot cycle inevitably causes a heat leakage~\cite{Schmiedl_2007,PhysRevE.97.022131,PhysRevE.101.032129}.
In the overdamped dynamics, the inertial effect of the Brownian particle is disregarded, and the system is only described by its position.
Nevertheless, heat leakage is related to the kinetic energy of the particle, as seen below.
When the overdamped limit is considered in the underdamped dynamics,
the averaged kinetic energy of the Brownian particle is equal to $k_{B}T/2$ in the isothermal process with temperature $T$, where $k_{B}$ is the Boltzmann constant.
Then, after the instantaneous adiabatic processes in the above cycle, the kinetic energy relaxes toward the temperature of the subsequent isothermal process, and an additional heat proportional to the temperature difference flows.
This heat leakage decreases the efficiency of the cycle.
Thus, we must consider the underdamped dynamics to evaluate the effect of the heat leakage on the efficiency and power of the Brownian Carnot cycle with the instantaneous adiabatic processes.
%Therefore, the compatibility of the Carnot efficiency with finite power may be questioned when we consider the underdamped dynamics of the Brownian particle even in the overdamped limit.

In this paper, we demonstrate that it is possible to achieve the Carnot efficiency in the underdamped finite-power Brownian Carnot cycle by considering the vanishing limit of the relaxation times of both position and velocity in the small temperature-difference regime, where the heat leakage due to the instantaneous adiabatic processes can be negligible.
As shown below, $\eta_C-\eta$ in Eq.~(\ref{general trade-off relation}) is proportional to the entropy production.
We show that the above compatibility is made possible by the diverging constant $A$ in Eq.~(\ref{general trade-off relation}) and the vanishing entropy production, which can be expressed in terms of the two relaxation times of the system.

The rest of this paper is organized as follows.
In Sec.~\ref{model}, we introduce the Brownian particle trapped by the harmonic potential and describe it by the underdamped Langevin equation.
We also introduce the isothermal process and instantaneous adiabatic process in this section.
In Sec.~\ref{Carnot cycle}, we construct the Carnot cycle using the Brownian particle.
In Sec.~\ref{numerical simulation}, we present the results of numerical simulations of the underdamped Brownian Carnot cycle
when we vary the temperature difference and the relaxation times of the system.
From these results, we demonstrate that the efficiency of our cycle approaches the Carnot efficiency
while maintaining finite power as the relaxation times vanish in the small temperature-difference regime.
In Sec.~\ref{theoretical analysis}, 
%we show that the entropy production
%in the isothermal processes approaches zero
%with finite heat flux when we take the above limit.
%Based on the trade-off relation Eq.~(\ref{general trade-off relation}),
we explain the results of the numerical simulations in Sec.~\ref{numerical simulation} based on the trade-off relation in Eq.~(\ref{general trade-off relation}).
Section~\ref{summary and discussion} presents the summary and discussion.

% Put \label in argument of \section for cross-referencing

\section{Model}
\label{model}
\subsection{Underdamped system}
\label{underdamped system}
We consider a Brownian particle in the surrounding medium with a temperature $T$.
When the particle is trapped in the harmonic potential
\begin{equation}
  \label{harmonic potential}
  V(x,t)=\frac{1}{2}\lambda(t)x^2,
\end{equation}
the dynamics of the particle is described by the underdamped Langevin equation
\begin{align}
  \label{Langevin equation of x}
  \dot{x} =& v,\\
  \label{Langevin equation of v}
  m \dot{v} =& -\gamma v - \lambda x +\sqrt{2\gamma k_{B} T} \xi,
\end{align}
where, $x$, $v$, and $m$ are the position,
velocity, and mass of the particle, respectively.
The dot denotes the time derivative or a quantity per unit time.
We use $\gamma$ as the constant friction coefficient independent of $T$
and set the Boltzmann constant $k_{B} = 1$ for simplicity.
The stiffness $\lambda(t)$ of the harmonic potential changes over time.
The Gaussian white noise $\xi(t)$ satisfies $\langle \xi(t)\rangle = 0$
and $\langle \xi(t)\xi(t')\rangle = \delta(t-t')$,
where $\langle \cdots \rangle$ denotes statistical average.
In this system, the relaxation times of the position $\tau_{x}$ and velocity $\tau_{v}$ are defined as follows:
\begin{align}
  \label{relaxation time of x}
  \tau_{x}(t) \equiv& \frac{\gamma}{\lambda(t)},\\
  \label{relaxation time of v}
  \tau_{v} \equiv& \frac{m}{\gamma},
\end{align}
where $\tau_{x}(t)$ depends on the time through the stiffness $\lambda(t)$.
We introduce the distribution function $p(x,v,t)$ to describe the state of the system at time $t$.
The time evolution of $p(x,v,t)$ can be described by
the Kramers equation~\cite{Risken_2nd_book} corresponding to
Eqs.~(\ref{Langevin equation of x}) and (\ref{Langevin equation of v}),
\begin{equation}
  \label{Fokker-Planck equation}
  \begin{split}
    \frac{\partial}{\partial t}p(x,v,t)
    =& -\frac{\partial}{\partial x}(v p(x,v,t))\\
    &+ \frac{\partial}{\partial v}\left[
      \frac{\gamma}{m}v + \frac{\lambda}{m}x
      + \frac{\gamma T}{m^2}\frac{\partial}{\partial v}
      \right]p(x,v,t)\\
      =& -\frac{\partial}{\partial x} j_{x}(x,v,t)
    -\frac{\partial}{\partial v} j_{v}(x,v,t),
  \end{split}
\end{equation}
where $j_{x}(x,v,t)$ and $j_{v}(x,v,t)$ are the probability currents defined as follows:
\begin{align}
  \label{the probability current of x}
  j_{x}(x,v,t) \equiv& v p(x,v,t),\\
  \label{the probability current of v}
  j_{v}(x,v,t) \equiv& -\left[\frac{\gamma}{m}v + \frac{\lambda}{m}x
    + \frac{\gamma T}{m^2}\frac{\partial}{\partial v}\right]p(x,v,t).
\end{align}
Here, we define the three variables
$\sigma_{x}(t) \equiv \langle x^2 \rangle$,
$\sigma_{v}(t) \equiv \langle v^2 \rangle$,
and $\sigma_{xv}(t) \equiv \langle xv \rangle$.
%When we use Eqs.~(\ref{Langevin equation of x}) and (\ref{Langevin equation of v}),
By using Eq.~(\ref{Fokker-Planck equation}),
we can derive the following equations: 
\begin{align}
  \label{equation of sigma_x}
  \dot{\sigma}_{x} =& 2\sigma_{xv},\\
  \label{equation of sigma_v}
  \dot{\sigma}_{v} =& \frac{2 \gamma T}{m^2} - \frac{2 \gamma}{m}\sigma_{v}
  - \frac{2 \lambda}{m}\sigma_{xv},\\
    \label{equation of sigma_xv}
  \dot{\sigma}_{xv} =& \sigma_{v} - \frac{\lambda}{m}\sigma_{x} - \frac{ \gamma}{m}\sigma_{xv}
\end{align}
describing the time evolution of $\sigma_{x}$, $\sigma_{v}$, and $\sigma_{xv}$~\cite{Dechant_2017}.
Below, we assume that the probability distribution $p(x,v,t)$ is a Gaussian distribution:
\begin{equation}
  \label{Gaussian distribution}
  \begin{split}
    p(x,v,t) =& \frac{1}{\sqrt{4 \pi^2(\sigma_{x}\sigma_{v}-\sigma_{xv}^2)}}\\
    &\times \exp \left\{ -\frac{\sigma_{x}v^2 + \sigma_{v}x^2 - 2\sigma_{xv} xv}
    {2(\sigma_{x}\sigma_{v}-\sigma_{xv}^2)}\right\}.
  \end{split}
\end{equation}
Thus, the state of the Brownian particle can only be described by the above three variables.
In this model, the internal energy $E(t)$ and entropy $S(t)$ of the Brownian particle are defined as follows:
\begin{align}
  \label{definition of internal energy}
    E(t) \equiv& \int^{\infty}_{-\infty} dx \int^{\infty}_{-\infty} dv\ p(x,v,t)\ \left[ \frac{1}{2}mv^2 + \frac{1}{2}\lambda(t) x^2\right] \nonumber\\
    &= \frac{1}{2}m\sigma_{v}(t) + \frac{1}{2}\lambda(t) \sigma_{x}(t),\\
  \label{definition of entropy}
    S(t) \equiv& -\int^{\infty}_{-\infty} dx\int^{\infty}_{-\infty} dv\ p(x,v,t)\ln p(x,v,t) \nonumber\\
    =&\frac{1}{2}\ln(\sigma_{x}(t)\sigma_{v}(t)-\sigma_{xv}^{2}(t)) + \ln(2\pi) + 1.
\end{align}

\subsection{Isothermal process}
\label{Isothermal process}
We define the heat and work during a time interval $t_{i} < t < t_{f}$
in an isothermal process.
In this process, the Brownian particle interacts with the heat bath at a constant temperature $T$.
We assume that the stiffness $\lambda(t)$ changes smoothly in this process.
The heat flux $\dot{Q} $ flowing from the heat bath to the Brownian particle
is defined as the statistical average of
the work performed by the force from the heat bath to the Brownian particle
(see Chap.~4 of Ref.~\cite{Sekimoto_book}),
\begin{equation}
  \label{definition of the heat flux}
  \dot{Q}(t) \equiv \left\langle \left(-\gamma v +\sqrt{2\gamma T} \xi(t)\right)\circ v\right\rangle,
\end{equation}
where $\circ$ represents the Stratonovich-type product.
Using Eqs.~(\ref{Langevin equation of x}) and (\ref{Langevin equation of v}),
we derive the heat flux $\dot{Q}(t)$ as follows:
\begin{equation}
  \label{heat flux expressed by variables}
  \dot{Q}(t)
  = \frac{1}{2}\lambda(t)\dot{\sigma}_{x}(t)+\frac{1}{2}m\dot{\sigma}_{v}(t).
\end{equation}
Thus, we obtain the heat $Q$ flowing in this interval as
\begin{equation}
  \label{heat expressed by variables}
  \begin{split}
    Q =& \int^{t_{f}}_{t_{i}}dt\ \left(\frac{1}{2}\lambda\dot{\sigma}_{x}\right)
    +\int^{t_{f}}_{t_{i}}dt\ \left(\frac{1}{2}m\dot{\sigma}_{v}\right)\\
    %=&\int^{\sigma_{x}(t_{f})}_{\sigma_{x}(t_{i})}\frac{1}{2}\lambda d\sigma_{x}+ \frac{1}{2}m\int^{\sigma_{v}(t_{f})}_{\sigma_{v}(t_{i})}d\sigma_{v}\\
    =& Q^{o} +\Delta K,
  \end{split}
\end{equation}
where
\begin{align}
  \label{heat derived from potential energy}
  Q^{o} \equiv& \int^{t_{f}}_{t_{i}}dt\ \left(\frac{1}{2}\lambda\dot{\sigma}_{x}\right),\\
  \label{difference of kinetic energy}
  \Delta K \equiv& \frac{1}{2}m\sigma_{v}(t_{f}) - \frac{1}{2}m\sigma_{v}(t_{i}).
\end{align}
Here, $Q^{o}$ represents the heat related to the potential change, and $\Delta K$ is the difference between the initial and final (averaged) kinetic energies of the Brownian particle.
In the overdamped system~\cite{Schmiedl_2007},
$Q^{o}$ is regarded as the heat instead of $Q$ in Eq.~(\ref{heat expressed by variables}).
However, in the underdamped system under consideration,
the heat also includes the kinetic part $\Delta K$.
%we does not treat $Q^{o}$ as the heat alone.
%As we showed above, the heat can be divided into two parts,
%the kinetic part $\Delta K$ and the potential part $Q^{o}$.

The output work during this interval is defined as follows:
\begin{equation}
  \label{definition of work in isothermal process}
  \begin{split}
    W  &\equiv
    -\int^{t_{f}}_{t_{i}}dt \int^{\infty}_{-\infty} dx \int^{\infty}_{-\infty} dv\ p(x,v,t) \frac{\partial V(x,t)}{\partial t} \\
    &= -\frac{1}{2}\int^{t_{f}}_{t_{i}}dt\ \dot{\lambda} \sigma_{x}\\
    &= Q - \Delta E ,
  \end{split}
\end{equation}
where we used Eqs.~(\ref{definition of internal energy})
and (\ref{heat expressed by variables}) for the derivation from the middle to the last equality,
and defined $\Delta E  \equiv E(t_{f}) - E(t_{i})$.
The last equality in Eq.~(\ref{definition of work in isothermal process}) represents the first law of thermodynamics.

\subsection{Instantaneous adiabatic process}
\label{Instantaneous adiabatic process}
As an adiabatic process connecting the end of the isothermal process with temperature $T_{1}$ to the beginning of the next isothermal process with temperature $T_{2}$, we use instantaneous changes in the potential and heat bath at $t=t_{0}$, which we regard as the final time of the isothermal process with temperature $T_{1}$~\cite{Schmiedl_2007}.
In this process, the stiffness $\lambda(t)$ jumps from $\lambda_{1}$
to $\lambda_{2}$,
and we instantaneously switch the temperature of
the heat bath from $T_{1}$ to $T_{2}$, maintaining the probability distribution unchanged.
Because this process is instantaneous, no heat exchange occurs,
and the output work $W^{ad}_{1 \to 2}$ is equal to the negative value of the internal energy change $\Delta E^{ad}_{1 \to 2}$ due to the first law of thermodynamics as
\begin{equation}
  \label{work of adiabatic}
  W^{ad}_{1 \to 2}= -\Delta E^{ad}_{1 \to 2}
  = -\frac{1}{2}(\lambda_{2} - \lambda_{1})\sigma_{x}(t_0).
\end{equation}

%There can be a large difference between the heat $Q^{o}$ in overdamped regime
%and that in underdamped regime $Q$ in Eq.~(\ref{heat expressed by the variables})
%\cite{Dechant_2017}.

\section{Carnot cycle}
\label{Carnot cycle}
\begin{figure}[t]
  \includegraphics[width=6.1cm]{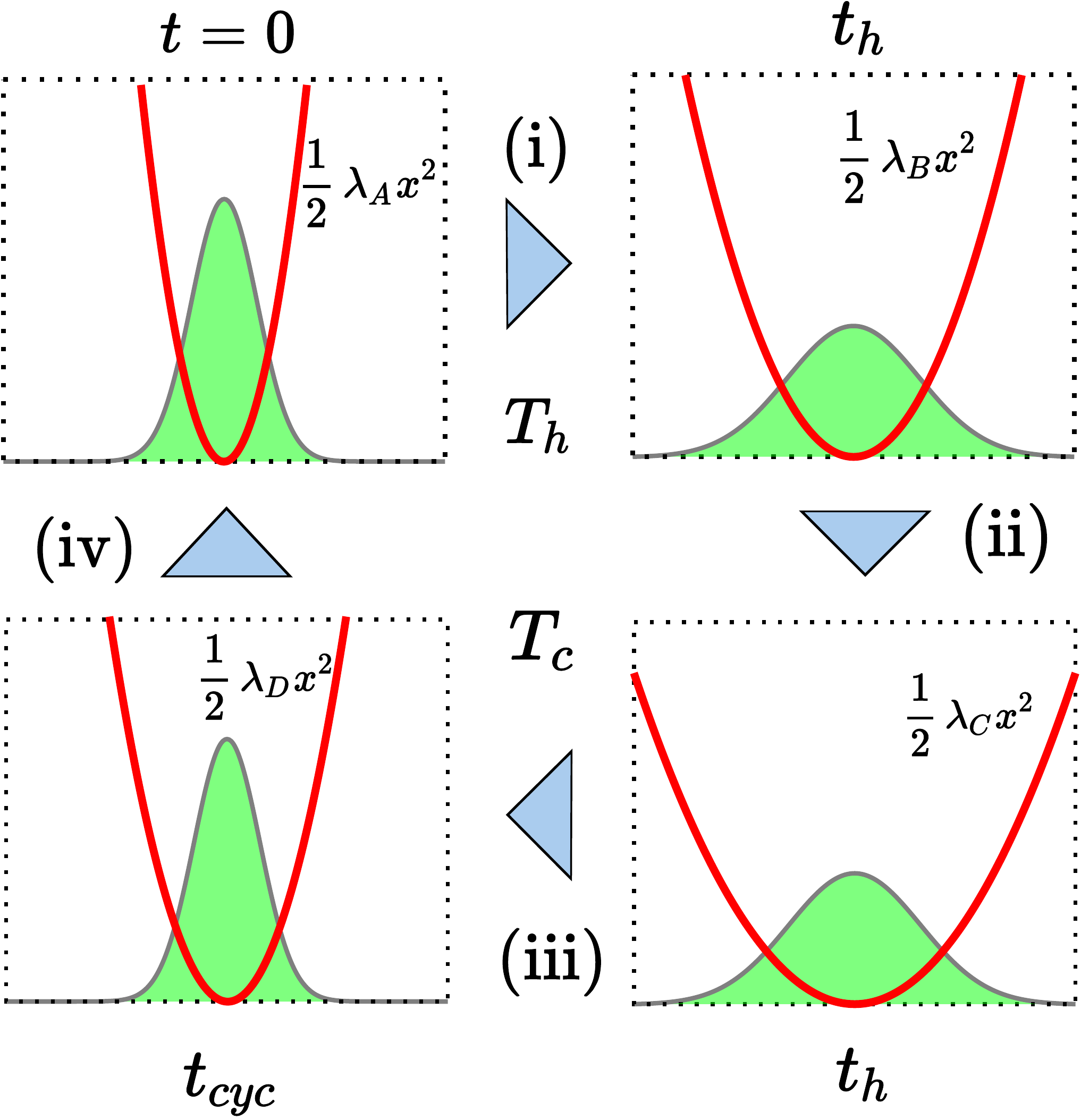}
  \caption{
    Schematic illustration of the Brownian Carnot cycle.
    In each box, the bottom horizontal line denotes the position coordinate $x$, and the boundary curve of the green filled area denotes the probability distribution of $x$. The red solid line corresponds to the harmonic potential.
    This cycle is composed of
    (i) hot isothermal process,
    (ii) instantaneous adiabatic process,
    (iii) cold isothermal process, and
    (iv) instantaneous adiabatic process.
  }
  \label{fig:schematic illustration}
\end{figure}
We construct a Carnot cycle
operating between the two heat baths with the temperatures
$T_h$ and $T_c$ (see Fig.~\ref{fig:schematic illustration}) 
by combining the isothermal processes and the instantaneous adiabatic processes introduced in Sec.~\ref{model}.

First, we define a protocol of a finite-time Carnot cycle with stiffness $\lambda(t)$ as follows: 
The hot isothermal process with temperature $T_{h}$ lasts for $0<t<t_{h}$, and
the stiffness $\lambda$ varies from $\lambda_{A}$ to $\lambda_{B}$
[Fig.~\ref{fig:schematic illustration}(i)].  
In the following instantaneous adiabatic process,
we switch the stiffness from $\lambda_{B}$ to $\lambda_{C}$ and the temperature of the heat bath from $T_{h}$ to $T_{c}$ at $t=t_{h}$, 
[Fig.~\ref{fig:schematic illustration}(ii)].
The cold isothermal process with temperature $T_{c}$ lasts for $t_{h}<t<t_{h}+t_{c}$,
and the stiffness $\lambda$ varies from $\lambda_{C}$ to $\lambda_{D}$
[Fig.~\ref{fig:schematic illustration}(iii)].
%Below, we use $t_{cyc}\equiv t_{h}+t_{c}$ as the cycle time.
In the last instantaneous adiabatic process,
we switch the stiffness from $\lambda_{D}$ to $\lambda_{A}$
and the temperature of the heat bath from $T_{c}$ to $T_{h}$ at $t=t_{cyc}$, Fig.~\ref{fig:schematic illustration}(iv), where $t_{cyc}\equiv t_{h}+t_{c}$ is the cycle time, which is assumed nonzero.
The final state of the Brownian particle in the cold (hot) isothermal process should agree with the initial state in the hot (cold) isothermal process.

We assume that the stiffness $\lambda(t)$ can be expressed as follows:
\begin{equation}
  \label{assumption of the protocol}  
  \lambda(t)=\Lambda(s) \hspace{3mm} \left(s \equiv \frac{t}{t_{cyc}}\right),
\end{equation}
using the scaling function 
$\Lambda(s)$ $(0\le s\le 1)$.
Under this assumption, we can change the time scale of the protocol maintaining the protocol form unchanged, by selecting another value of $t_{cyc}$.
We also assume that $t_{h}/t_{cyc}$ and $t_{c}/t_{cyc}$ are finitely fixed for any value of $t_{cyc}$.
Furthermore, we assume that $\lambda(t_f)/\lambda(t_i)$ is finite at any time $t_i$ and $t_f$, where they are in the same isothermal process.
We use this assumption to show that the heat flux after the relaxation at the beginning of the isothermal processes is noninfinite in the Appendix.
Note that the word ``finite" may situationally be used considering two meanings, ``nonzero" (e.g., ``finite power") or ``noninfinite" (e.g., ``finite time").
In this paper, however, we refer to ``nonzero and noninfinite" by ``finite" except for the two examples above.

To consider the quasistatic Carnot cycle
corresponding to the above finite-time Carnot cycle,
we must consider the limit of $t_{cyc}\to \infty$
and use the stiffness $\lambda^{qs}(t)$
related to the finite-time stiffness through Eq.~(\ref{assumption of the protocol}).
Here, the index $qs$ of $X^{qs}$ denotes the physical quantity $X$ evaluated
in the quasistatic limit.

\subsection{Quasistatic Carnot cycle: Quasistatic efficiency}
\label{Quasistatic Carnot cycle}
We formulate the efficiency of the quasistatic Carnot cycle.
To this end, we need to quantify the heat leakage caused by the adiabatic process. 
As the adiabatic processes are instantaneous, the initial distributions of the quasistatic isothermal processes do not agree 
with the equilibrium distributions at the temperature of the heat bath.
%First, we consider the quasistatic isothermal processes.
Thus, a relaxation at the beginning of the isothermal processes exists, and in general, the relaxation is irreversible.
%At the beginning of each isothermal process,
%there is a relaxation due to the instantaneous adiabatic processes.
%In general, the relaxation is irreversible.
After the relaxation in the quasistatic isothermal process with temperature $T$,
the time derivative of the variables satisfies
\begin{equation}
  \dot{\sigma}_{x}^{qs}(t) = 0,
  \quad
  \dot{\sigma}_{v}^{qs}(t) = 0,
  \quad
  \dot{\sigma}_{xv}^{qs}(t) = 0.
\end{equation}
Subsequently, from Eqs.~(\ref{equation of sigma_x})--(\ref{equation of sigma_xv}),
we obtain those values as follows:
\begin{equation}
  \label{variable in the quasistatic limit}
  \sigma_{x}^{qs}(t) = \frac{T}{\lambda^{qs}(t)},
  \quad
  \sigma_{v}^{qs}(t) = \frac{T}{m},
  \quad
  \sigma_{xv}^{qs}(t) = 0,
\end{equation}
and the distribution in Eq.~(\ref{Gaussian distribution}) in the quasistatic limit
agrees with the Boltzmann distribution
\begin{equation}
  \label{Boltzmann distribution}
  p^{qs}(x,v,t)=\sqrt{\frac{m\lambda^{qs}(t) }{4 \pi^{2} T^2}}
  \exp \left\{ -\frac{\lambda^{qs}(t) x^2+m v^2}{2T}\right\}.
\end{equation}
After the relaxation in each quasistatic isothermal process,
the system is in equilibrium with the heat bath and satisfies Eq.~(\ref{variable in the quasistatic limit}).
Using Eqs.~(\ref{definition of entropy}) and (\ref{variable in the quasistatic limit}), 
we derive the quasistatic entropy as follows:
\begin{align}
\label{entropy in the equilibrium}
    S^{qs}(t)=& \frac{1}{2}\ln\sigma_x^{qs}(t)
    +\frac{1}{2}\ln\sigma_v^{qs}(t)
    +\ln(2\pi)+1\nonumber\\
    =&\frac{1}{2}\ln\left(\frac{T}{\lambda(t)}\right)
    +\frac{1}{2}\ln\left(\frac{T}{m}\right)
    +\ln(2\pi)+1.
\end{align}

As mentioned above, the quasistatic isothermal processes are
composed of the relaxation part and the part after the relaxation.
Because the instantaneous adiabatic process [Fig.~\ref{fig:schematic illustration}(iv)] just before the quasistatic hot isothermal process [Fig.~\ref{fig:schematic illustration}(i)] does not change the probability distribution, the initial distribution agrees with the final distribution in the quasistatic cold isothermal process.
Thus, the variables $\sigma_{x}^{qs}$, $\sigma_{v}^{qs}$, and $\sigma_{xv}^{qs}$ begin the quasistatic hot isothermal process with the following values: 
\begin{equation}
  \label{equilibrium state D}
  \sigma_{x}^{qs}=\frac{T_{c}}{\lambda_{D}^{qs}},
  \quad
  \sigma_{v}^{qs}=\frac{T_{c}}{m},
  \quad
  \sigma_{xv}^{qs}=0,
\end{equation}
where we used Eq.~(\ref{variable in the quasistatic limit}).
In the relaxation at the beginning of this process,
the stiffness almost remains $\lambda_{A}^{qs}$
[see Eq.~(\ref{stiffness in the relaxation}) in the Appendix],
and the variables relax to
\begin{equation}
  \label{equilibrium state A}
  \sigma_{x}^{qs}=\frac{T_{h}}{\lambda_{A}^{qs}},
  \quad
  \sigma_{v}^{qs}=\frac{T_{h}}{m},
  \quad
  \sigma_{xv}^{qs}=0,
\end{equation}
owing to Eq.~(\ref{variable in the quasistatic limit}).

From Eqs.~(\ref{equilibrium state D}) and (\ref{equilibrium state A}),
the kinetic energy is $m\sigma_{v}/2=T_c/2$ in the initial state and
changes to $T_h/2$ during the relaxation.
The kinetic energy remains $T_h/2$ after the relaxation
because the system is in equilibrium with the heat bath at temperature $T_{h}$ during the quasistatic hot isothermal process.
Thus, a change in the kinetic energy in Eq.~(\ref{difference of kinetic energy})
in the quasistatic hot isothermal process is given by 
\begin{equation}
    \label{Delta K qs h}
    \Delta K_{h}^{qs}=\frac{\Delta T}{2},
\end{equation}
where $\Delta T \equiv T_{h}-T_{c}$.
We can also derive the heat related to
the potential change during the relaxation
$Q^{rel,o,qs}_{h}$ as follows.
As the stiffness remains $\lambda_{A}^{qs}$ during
the relaxation,
$Q^{rel,o,qs}_{h}$ is derived as
\begin{align}
  %\label{heat in the relaxation in the hot isothermal process}
  \label{heat rel o qs h}
    Q^{rel,o,qs}_{h}
    =&\int^{T_{h}/\lambda_{A}^{qs}}_{T_{c}/\lambda_{D}^{qs}}
    \frac{1}{2}\lambda_{A}^{qs}d\sigma_{x}\nonumber\\
    =& \frac{1}{2}\lambda^{qs}_{A}
    \left( \frac{T_{h}}{\lambda^{qs}_{A}}-\frac{T_{c}}{\lambda^{qs}_{D}}\right),
\end{align}
using Eq.~(\ref{heat derived from potential energy}).
The entropy change of the Brownian particle in this relaxation is given by
\begin{equation}
  \label{the entropy production in the quasistatic hot isothermal process}
  \Delta S_h^{rel,qs}\equiv\frac{1}{2}\ln\left(\frac{T_{h}}{\lambda_{A}^{qs}}\frac{\lambda_{D}^{qs}}{T_c}\right)+\frac{1}{2}\ln\left(\frac{T_{h}}{T_c}\right),
\end{equation}
%where we have used Eqs.~(\ref{entropy in the equilibrium}), (\ref{equilibrium state D}), and (\ref{equilibrium state A}).
where we used Eqs.~(\ref{entropy in the equilibrium})--(\ref{equilibrium state A}).

After the relaxation in the quasistatic hot isothermal process, the probability distribution maintains
the Boltzmann distribution in Eq.~(\ref{Boltzmann distribution}) with $T=T_h$,
and $\sigma_v$ does not change.
Therefore, the final state of the process should satisfy
\begin{equation}
  \label{equilibrium state B}
  \sigma_{x}^{qs}=\frac{T_{h}}{\lambda_{B}^{qs}}
  \quad
  \sigma_{v}^{qs}=\frac{T_{h}}{m},
  \quad
  \sigma_{xv}^{qs}=0,
\end{equation}
where we used Eq.~(\ref{variable in the quasistatic limit}).
Because the second term on the right-hand side of Eq.~(\ref{entropy in the equilibrium}) does not change in the quasistatic hot isothermal process, we derive the entropy change $\Delta S_{h}^{iso,qs}$
after the relaxation in this process as follows:
\begin{equation}
  \label{the entropy change in the quasistatic hot isothermal process}
  \Delta S_{h}^{iso,qs} \equiv \frac{1}{2}\ln
  \left( \frac{\lambda_{A}^{qs}}{\lambda_{B}^{qs}}\right).
\end{equation}
Note that the quantities with the index ``$iso$" do not include the contribution from the relaxation.
Thus, the heat supplied to the Brownian particle after the relaxation
in this process is given by
\begin{equation}
  \begin{split}
    \label{heat in the quasistatic hot isothermal process after the relaxation}
    %Q^{iso,o,qs}_{h}\equiv&
    T_{h}\Delta S_{h}^{iso,qs}
    =& \frac{T_{h}}{2}\ln \left(
    \frac{\lambda_{A}^{qs}}{\lambda_{B}^{qs}}
    \right).    
  \end{split}
\end{equation}
%by using Eq.~(\ref{the entropy change in the quasistatic hot isothermal process}).
The heat related to the potential change in the quasistatic hot isothermal process is 
\begin{eqnarray}
\label{Q o qs h}
  Q^{o,qs}_{h}=T_{h}\Delta S_{h}^{iso,qs}+Q^{rel,o,qs}_{h}.
\end{eqnarray}
Therefore, by using Eq.~(\ref{Delta K qs h}), the heat flowing
in the quasistatic hot isothermal process is given by
\begin{align}
  \label{heat in the quasistatic hot isothermal process}
      Q_{h}^{qs} = &Q^{o,qs}_{h}+ \Delta K_{h}^{qs} \nonumber\\
    =&T_{h}\Delta S_{h}^{iso,qs} + Q^{rel,o,qs}_{h}+ \frac{1}{2}\Delta T, \\
    =&T_{h}\Delta S_{h}^{iso,qs} + Q^{rel,qs}_{h},\nonumber
\end{align}
where $Q^{rel,qs}_{h}$ denotes the heat flowing during the relaxation at the beginning of this process, as
\begin{equation}
  \label{heat rel qs h}
  Q^{rel,qs}_{h} \equiv Q^{rel,o,qs}_{h}+\frac{1}{2}\Delta T.
\end{equation}
%and (\ref{heat in the relaxation in the hot isothermal process}).
From Eq.~(\ref{definition of work in isothermal process}), the work in this process is given by
\begin{equation}
  \label{quasistatic work in the hot isothermal process}
  W^{qs}_{h}=Q_{h}^{qs}-\Delta E^{qs}_{h},
\end{equation}
where $\Delta E^{qs}_{h}$ represents the internal energy change in this process.

After the instantaneous adiabatic process [Fig.~\ref{fig:schematic illustration}(ii)], the quasistatic cold isothermal process [Fig.~\ref{fig:schematic illustration}(iii)] begins with the variables in Eq.~(\ref{equilibrium state B}), and the variables relax to
\begin{equation}
  \label{equilibrium state C}
  \sigma_{x}^{qs}=\frac{T_{c}}{\lambda_{C}^{qs}}
  \quad
  \sigma_{v}^{qs}=\frac{T_{c}}{m},
  \quad
  \sigma_{xv}^{qs}=0,
\end{equation}
where we used Eq.~(\ref{variable in the quasistatic limit}).
Similar to the quasistatic hot isothermal process, the change in the kinetic energy in Eq.~(\ref{difference of kinetic energy}) satisfies
\begin{equation}
\label{Delta K qs c}
\Delta K_c^{qs}=-\frac{\Delta T}{2}.
\end{equation}
We also define the heat related to the potential change during the relaxation in the quasistatic cold isothermal process as
\begin{equation}
  \label{heat rel o qs c}
  Q^{rel,o,qs}_{c}\equiv \frac{1}{2}\lambda^{qs}_{C}\left( \frac{T_{c}}{\lambda^{qs}_{C}}-\frac{T_{h}}{\lambda^{qs}_{B}}\right).
\end{equation}
Then, the flowing heat and the entropy change of the particle during this relaxation are given by
\begin{align}
 \label{heat rel qs c}
  Q^{rel,qs}_{c} \equiv& %Q^{rel,o,qs}
   \frac{1}{2}\lambda^{qs}_{C}\left( \frac{T_{c}}{\lambda^{qs}_{C}}-\frac{T_{h}}{\lambda^{qs}_{B}}\right)
  -\frac{1}{2}\Delta T,\\
  \label{the entropy production in the quasistatic cold isothermal process}
  \Delta S_c^{rel,qs}\equiv&\frac{1}{2}\ln\left(\frac{T_{c}}{\lambda_{C}^{qs}}\frac{\lambda_{B}^{qs}}{T_{h}}\right)+\frac{1}{2}\ln\left(\frac{T_{c}}{T_{h}}\right),
\end{align}
similarly to Eqs.~(\ref{the entropy production in the quasistatic hot isothermal process}) and (\ref{heat rel qs h}), where we used Eqs.~(\ref{heat derived from potential energy}),
(\ref{entropy in the equilibrium}),
(\ref{equilibrium state B}), and (\ref{equilibrium state C})--(\ref{heat rel o qs c}).

After the relaxation, the variables change to the state in Eq.~(\ref{equilibrium state D}).
Then, the entropy change after the relaxation in the quasistatic cold isothermal process is given by
\begin{equation}
  \label{the entropy change in the quasistatic cold isothermal process}  
  \Delta S^{iso,qs}_{c}\equiv \frac{1}{2}\ln\left(\frac{\lambda^{qs}_{C}}{\lambda^{qs}_{D}}\right).
\end{equation}
The heat related to the potential change in the quasistatic cold isothermal process is 
\begin{eqnarray}
\label{Q o qs c}
  Q^{o,qs}_{c}=T_{c}\Delta S_{c}^{iso,qs}+Q^{rel,o,qs}_{h},
\end{eqnarray}
where we used Eqs.~(\ref{heat rel o qs c}) and (\ref{the entropy change in the quasistatic cold isothermal process}).
Thus, the heat flowing in the quasistatic cold isothermal process is given by
\begin{align}
  \label{heat in the quasistatic cold isothermal process}
    Q_{c}^{qs} = &Q^{o,qs}_{c}+ \Delta K_{c}^{qs}\nonumber\\
    %=&T_{c}\Delta S^{qs}_{c} +Q^{rel,o,qs}_{c} - \frac{1}{2}\Delta T \\
    =&T_{c}\Delta S_{c}^{iso,qs} + Q^{rel,qs}_{c},
\end{align}
where we used Eqs.~(\ref{Delta K qs c})--(\ref{Q o qs c}).
From Eq.~(\ref{definition of work in isothermal process}), the work in this process is given by
\begin{equation}
  \label{quasistatic work in the cold isothermal process}
  W^{qs}_{c}=Q_{c}^{qs}-\Delta E^{qs}_{c},
\end{equation}
where $\Delta E^{qs}_{c}$ is the internal energy change in this process.
After the quasistatic cold isothermal process [Fig.~\ref{fig:schematic illustration}(iii)], the system proceeds to the 
instantaneous adiabatic process [Fig.~\ref{fig:schematic illustration}(iv)] and returns to
the initial state of the quasistatic hot isothermal process.

%From Eqs.~(\ref{work of adiabatic}), (\ref{equilibrium state D}), (\ref{equilibrium state A}), (\ref{equilibrium state B}), and (\ref{equilibrium state C}), the output works in the instantaneous adiabatic processes are given by
%\begin{equation}
%\label{work of adiabatic process from h to c in quasistatic cycle}
%\begin{split}
%W^{ad,qs}_{h\to c} =-\Delta E^{ad,qs}_{h\to c} =& -\frac{1}{2}(\lambda_{C}^{qs} - \lambda_{B}^{qs})\frac{T_h}{\lambda^{qs}_{B}},
%\end{split}
%\end{equation}  
%\begin{equation}
%\label{work of adiabatic process from c to h in quasistatic cycle}
%\begin{split}
%W^{ad,qs}_{c\to h} = -\Delta E^{ad,qs}_{c\to h}=&-\frac{1}{2}(\lambda_{A}^{qs} - \lambda_{D}^{qs})\frac{T_c}{\lambda^{qs}_{D}}.
%\end{split}
%\end{equation}

Subsequently, we consider the efficiency of the quasistatic Carnot cycle.
As the cycle closes, the entropy change in the particle per cycle vanishes as
\begin{equation}
\label{total entropy change of quasistatic Carnot cycle}
    \Delta S_h^{rel,qs}+\Delta S_h^{iso,qs}+\Delta S_c^{rel,qs}+\Delta S_c^{iso,qs}=0,
\end{equation}
where we used  Eqs.~(\ref{the entropy production in the quasistatic hot isothermal process}),
(\ref{the entropy change in the quasistatic hot isothermal process}), 
(\ref{the entropy production in the quasistatic cold isothermal process}), and 
(\ref{the entropy change in the quasistatic cold isothermal process}).
Because the internal energy change in the particle per cycle vanishes, we derive the work per cycle from the first law of thermodynamics as
\begin{equation}
  \label{quasistatic work}
  \begin{split}
    W^{qs} 
    %=& W^{iso,qs}_{h} + W^{iso,qs}_{c} + W^{ad,qs}_{h\to c}+W^{ad,qs}_{c\to h}\\
    =&Q_{h}^{qs}+Q_{c}^{qs},
  \end{split}
\end{equation}
using Eqs.~(\ref{quasistatic work in the hot isothermal process}) and
(\ref{quasistatic work in the cold isothermal process}).
In our quasistatic cycle, the entropy production per cycle $\Sigma^{qs}$, by which we imply the total entropy production per cycle including the particle and heat baths, is obtained as follows:
\begin{equation}
\label{entropy production in quasistatic cycle}
    \Sigma^{qs}\equiv -\frac{Q_h^{qs}}{T_h}-\frac{Q_c^{qs}}{T_c}.
\end{equation}
Because an entropy change in the particle per cycle vanishes, as seen from Eq.~(\ref{total entropy change of quasistatic Carnot cycle}), the entropy production per cycle $\Sigma^{qs}$ is expressed only by the entropy change of the heat baths.
Using Eqs.~(\ref{heat in the quasistatic hot isothermal process}), (\ref{quasistatic work}), and (\ref{entropy production in quasistatic cycle}), we can derive the quasistatic efficiency as
\begin{equation}
  \label{quasistatic efficiency}
  \eta^{qs}\equiv \frac{W^{qs}}{Q^{qs}_{h}}=\eta_C - \frac{T_c\Sigma^{qs}}{Q_h^{qs}}.
\end{equation}
From Eq.~(\ref{quasistatic efficiency}), $\Sigma^{qs}$ should vanish to obtain $\eta_C$.
Using Eqs.~(\ref{heat in the quasistatic hot isothermal process}), (\ref{heat in the quasistatic cold isothermal process}), and (\ref{total entropy change of quasistatic Carnot cycle}), we can rewrite $\Sigma^{qs}$ in Eq.~(\ref{entropy production in quasistatic cycle}) as
\begin{align}
\label{Sigma qs with lambda}
    \Sigma^{qs}=&
    -\frac{T_h\Delta S^{iso,qs}_h+Q^{rel,qs}_h}{T_h}-\frac{T_c\Delta S^{iso,qs}_c+Q^{rel,qs}_c}{T_c}\nonumber\\
    =&\Delta S^{rel,qs}_h-\frac{Q^{rel,qs}_h}{T_h}+\Delta S^{rel,qs}_c-\frac{Q^{rel,qs}_c}{T_c}\nonumber\\
    =&\frac{1}{2}\left(-\ln\left(\frac{T_c\lambda_{A}^{qs}}{T_h\lambda_{D}^{qs}}\right)+\frac{T_c\lambda^{qs}_{A}}{T_h\lambda^{qs}_{D}}-1\right)\\    
    &+\frac{1}{2}\left(-\ln\left(\frac{T_h\lambda_{C}^{qs}}{T_c\lambda_{B}^{qs}}\right)
    +\frac{T_h\lambda^{qs}_{C}}{T_c\lambda^{qs}_{B}}-1\right)+\frac{(\Delta T)^2}{2T_hT_c},\nonumber
\end{align}
where we used Eqs.~(\ref{the entropy production in the quasistatic hot isothermal process}), (\ref{heat rel qs h}), (\ref{heat rel qs c}), and (\ref{the entropy production in the quasistatic cold isothermal process}) at the last equality.
The first and second terms on the right-hand side of Eq.~(\ref{Sigma qs with lambda}), derived from $Q^{rel,o,qs}_{h,c}$ in Eqs.~(\ref{heat rel o qs h}) and (\ref{heat rel o qs c}) and the first  term of $\Delta S^{rel,qs}_{h,c}$ in Eqs.~(\ref{the entropy production in the quasistatic hot isothermal process}) and (\ref{the entropy production in the quasistatic cold isothermal process}), denote the entropy production related to the potential energy in the relaxation in the hot and cold isothermal processes, respectively.
The last term of Eq.~(\ref{Sigma qs with lambda}) comes from the heat related to the kinetic energy.
To achieve the Carnot efficiency, the entropy production should vanish, as shown in Eq.~(\ref{quasistatic efficiency}).
In the overdamped Brownian Carnot cycle with the instantaneous adiabatic process in previous studies~\cite{Schmiedl_2007,PhysRevE.96.062107,PhysRevLett.121.120601}, the Carnot efficiency is obtained  in the quasistatic limit.
In the overdamped cycle, $(\Delta T)^2/(2T_hT_c)$ in Eq.~(\ref{Sigma qs with lambda}) does not exist because $\sigma_v$ is not considered.
Thus, the entropy production in the overdamped cycle is given by
\begin{equation}
    \Sigma^{o,qs} \equiv f\left(\frac{T_c\lambda^{qs}_{A}}{T_h\lambda^{qs}_{D}}\right)+f\left(\frac{T_h\lambda^{qs}_{C}}{T_c\lambda^{qs}_{B}}\right),
\end{equation}
where $f$ is defined as
\begin{equation}
    f(u)\equiv -\ln u+u-1,
\end{equation}
where $f(u)$ is a downwardly convex function with the minimum value of $f(1)=0$.
Thus, for the entropy production $\Sigma^{o,qs}$ to vanish, the following condition is derived: 
\begin{equation}
  \label{condition of the adiabatic process to achieve the Carnot efficiency}
  \frac{T_{h}}{\lambda^{qs}_{A}}=\frac{T_{c}}{\lambda^{qs}_{D}},\hspace{5mm}
  \frac{T_{h}}{\lambda^{qs}_{B}}=\frac{T_{c}}{\lambda^{qs}_{C}}.
\end{equation}
This condition was adopted in the previous studies on the overdamped Brownian Carnot cycle ~\cite{Schmiedl_2007,PhysRevE.96.062107,PhysRevLett.121.120601}  in the quasistatic limit. 
We impose this condition on our underdamped cycle to reduce entropy production.
Then, we obtain
\begin{equation}
    \Delta S^{rel,qs}_h+\Delta S^{rel,qs}_c=0,
\end{equation}
using Eqs.~(\ref{the entropy production in the quasistatic hot isothermal process}) and
(\ref{the entropy production in the quasistatic cold isothermal process}).
Thus, from Eq.~(\ref{total entropy change of quasistatic Carnot cycle}),
we derive 
\begin{equation}
  \label{definition of entropy change}
  \Delta S^{iso,qs}_{h} = -\Delta S^{iso,qs}_{c} \equiv \Delta S^{qs}.
\end{equation}
In addition, because $Q^{rel,o,qs}_h$ in Eq.~(\ref{heat rel o qs h}) and $Q^{rel,o,qs}_c$ in (\ref{heat rel o qs c}) vanish, we obtain

\begin{align}
\label{heat with closed condition_1}
    &Q_h^{o,qs}=T_h\Delta S^{qs},
    \quad
    Q_c^{o,qs}=-T_c\Delta S^{qs},\\
    &Q_h^{rel,qs}=-Q_c^{rel,qs}=\frac{1}{2}\Delta T,
\end{align}
using Eqs.~(\ref{Q o qs h}), (\ref{heat rel qs h}), (\ref{heat rel qs c}), and (\ref{Q o qs c}).
The heat in Eqs.~(\ref{heat in the quasistatic hot isothermal process}) and 
(\ref{heat in the quasistatic cold isothermal process}) can also be rewritten as follows:
\begin{equation}
\label{heat with closed condition_2}
    Q_h^{qs}=T_h\Delta S^{qs} +\frac{1}{2}\Delta T,
    \quad
    Q_c^{qs}=-T_c\Delta S^{qs} -\frac{1}{2}\Delta T.
\end{equation}
Using Eqs.~(\ref{quasistatic work}) and (\ref{heat with closed condition_2}),
We can rewrite the work in Eq.~(\ref{quasistatic work}) and the efficiency in Eq.~(\ref{quasistatic efficiency}) as follows: 
\begin{align}
  \label{quasistatic work with adiabatic condition}
    W^{qs}
    =&\Delta T\Delta S^{qs},\\
  \label{quasistatic efficiency under the adiabatic condition}
  \eta^{qs}\equiv& \frac{W^{qs}}{Q^{qs}_{h}}
  = \frac{\Delta T\Delta S^{qs}}{T_{h}\Delta S^{qs} +\frac{1}{2}\Delta T}<\eta_{C}.
\end{align}
Despite considering the quasistatic limit of our Carnot cycle,
however, the quasistatic efficiency $\eta^{qs}$ is smaller than
the Carnot efficiency because of the heat leakage $\Delta T/2$ in the denominator in Eq.~(\ref{quasistatic efficiency under the adiabatic condition}),
which is derived from a kinetic energy change in the particle due to the relaxation.

Here, we consider the small temperature-difference regime $\Delta T\to 0$ and
assume that $\Delta S^{qs} = O(1) >0$. 
Then, we obtain $\Delta T \Delta S^{qs} = O(\Delta T)$.
As the contribution of the heat leakage to $\eta^{qs}$
in Eq.~(\ref{quasistatic efficiency under the adiabatic condition}) can be of a higher order of $\Delta T$
in the small temperature-difference regime,
$\eta^{qs}$ is approximated by the Carnot efficiency as 
\begin{equation}
  \label{efficiency in the small temperature difference}
  \eta^{qs}= \frac{\Delta T\Delta S^{qs}}{T_{h}\Delta S^{qs}}+O[(\Delta T)^2]
  =\eta_{C}+O[(\Delta T)^2].
\end{equation}

\subsection{Finite-time Carnot cycle: Efficiency and power}
\label{finite-time Carnot cycle}
In the following, we formulate the efficiency and power of the finite-time Carnot cycle.
We assume that Eq.~(\ref{condition of the adiabatic process to achieve the Carnot efficiency}) is satisfied in the quasistatic limit of this cycle.
When we use the protocol in Eq.~(\ref{assumption of the protocol}), we obtain
\begin{equation}
  \label{edges of the protocol}
  \lambda_{i}^{qs}=\lambda_{i}\hspace{3mm}(i=A,B,C,D),
\end{equation}
and we can remove the index ``$qs$" in Eq.~(\ref{condition of the adiabatic process to achieve the Carnot efficiency}).
%In the quasistatic Carnot cycle, the work and heat are expressed by using $\Delta S$  in Eq.~(\ref{Delta S}).
In general, finite-time processes are irreversible, and the work and heat of the finite-time isothermal processes are different from those of quasistatic processes.
Thus, we express the work and heat in our finite-time cycle by using those in the quasistatic limit and the differences between the finite-time and quasistatic quantities.
Below, we mainly consider the finite-time Carnot cycle.
Thus, when we deal with a finite-time isothermal process or a finite-time cycle, we simply refer to them as an isothermal process or a cycle, respectively.
%To consider the entropy change of finite-time isothermal
%processes after the relaxation,
%we have to use the protocol which is different from
%that of the quasistatic one.
Using Eq.~(\ref{edges of the protocol}), we can rewrite the entropy changes in Eqs.~(\ref{the entropy change in the quasistatic hot isothermal process}) and (\ref{the entropy change in the quasistatic cold isothermal process}) in terms of the stiffness $\lambda(t)$ as
\begin{align}
  \label{quasistatic and finite-time entropy change in the hot isothermal process}
  \Delta S^{iso,qs}_{h}=&\frac{1}{2}\ln\left(\frac{\lambda_{A}^{qs}}{\lambda_{B}^{qs}}\right)
  =\frac{1}{2}\ln\left(\frac{\lambda_{A}}{\lambda_{B}}\right)
  =\Delta S^{qs},\\
  \label{quasistatic and finite-time entropy change in the cold isothermal process}
  \Delta S^{iso,qs}_{c}=&\frac{1}{2}\ln\left(\frac{\lambda_{C}^{qs}}{\lambda_{D}^{qs}}\right)
  =\frac{1}{2}\ln\left(\frac{\lambda_{C}}{\lambda_{D}}\right)
  =-\Delta S^{qs}.
\end{align}
From Eq.~(\ref{heat expressed by variables}), 
we derive the heat flowing from the hot heat bath to
the Brownian particle in the hot isothermal process as
\begin{equation}
  \label{supplied heat in the hot isothermal process}
  Q_{h} =  Q^{o}_{h} +\Delta K_{h} ,
\end{equation}
where
\begin{equation}
  \label{heat derived from the potential energy in the hot isothermal process}
  \begin{split}
    &Q^{o}_{h} = \frac{1}{2}\int^{t_{h}}_{0} dt\ \lambda\dot{\sigma}_{x},\\
    &\Delta K_{h}=\frac{1}{2}m \sigma_{v}(t_{h})-\frac{1}{2}m\sigma_{v}(0).
  \end{split}
\end{equation}
Note that $Q^{o}_{h}$ and $\Delta K_h$ become $Q^{o,qs}_{h}=T_h\Delta S^{qs}$ in Eq.~(\ref{heat with closed condition_1}) and $\Delta K^{qs}_h=\Delta T/2$ in Eq.~(\ref{Delta K qs h}), respectively, under the condition of Eq.~(\ref{condition of the adiabatic process to achieve the Carnot efficiency}) in the quasistatic limit, as discussed in Sec.~\ref{Quasistatic Carnot cycle}. 
Moreover, we find that $Q_h^o$ and $\Delta K_h$ differ from $T_h\Delta S^{qs}$ and $\Delta T/2$ because the process is not quasistatic.
Here, we define the irreversible work $W^{irr}_{h}$ to measure the difference between $Q_{h}^{o}$ and $T_{h}\Delta S^{qs}$ as
\begin{equation}
  \label{irreversible work in the hot isothermal process}
  \begin{split}
    W_{h}^{irr} %\equiv& T_{h}\Delta S_{h}^{qs} - Q_{h}^{o}\\
    \equiv &T_{h}\Delta S^{qs} - Q_{h}^{o}.
  \end{split}
\end{equation}
Then, the heat in the hot isothermal process in Eq.~(\ref{supplied heat in the hot isothermal process}) can be rewritten as follows:
\begin{equation}
  \label{the heat in the hot isothermal process}
  \begin{split}
    Q_{h} =& T_{h}\Delta S^{qs} - W_{h}^{irr} +\Delta K_{h},
  \end{split}
\end{equation}
using Eqs.~(\ref{supplied heat in the hot isothermal process}) 
and (\ref{irreversible work in the hot isothermal process}).
Moreover, using Eqs.~(\ref{definition of work in isothermal process})
and (\ref{the heat in the hot isothermal process}),
we obtain the output work in the hot isothermal process as
\begin{equation}
  \label{work in hot isothermal process}
  W_{h}
  =  T_{h}\Delta S^{qs} - W_{h}^{irr} + \Delta K_{h} - \Delta E_{h}, 
\end{equation}
where $\Delta E_{h}$
%$=E(t_h)-E(0)$ 
represents the internal energy change in this process.
The reason that we call $W^{irr}_{h}$ the irreversible work will be clarified later 
when we consider the output work per cycle.

The heat in Eq.~(\ref{heat expressed by variables}) in the cold isothermal process is given by
\begin{equation}
  \label{supplied heat in the cold isothermal process}
  Q_{c} =   Q^{o}_{c} +\Delta K_{c} ,
\end{equation}
where
\begin{align}
  \label{heat derived from the potential energy in the cold isothermal process}
  Q^{o}_{c} =& \frac{1}{2}\int^{t_{cyc}}_{t_{h}} dt\ \lambda\dot{\sigma}_{x},\\
    \Delta K_{c}=&\frac{1}{2}m\sigma_{v}(t_{cyc})-\frac{1}{2}m\sigma_{v}(t_{h})=-\Delta K_{h}.
     \label{kinetic energy change in the cold isothermal process}
\end{align}
Similar to $Q_{h}^{o}$ and $\Delta K_h$, $Q^{o}_{c}$ becomes $-T_c\Delta S^{qs}$ and $\Delta K_c$ becomes $-\Delta T/2$ under the condition of Eq.~(\ref{condition of the adiabatic process to achieve the Carnot efficiency}) in the quasistatic limit. 
In the same way as the hot isothermal process,
we can define the irreversible work $W^{irr}_{c}$  in this process
and rewrite the heat
in Eq.~(\ref{supplied heat in the cold isothermal process}) as follows:
\begin{align}
  \label{irreversible work in the cold isothermal process}
      W_{c}^{irr}
    \equiv& -T_{c}\Delta S^{qs} -Q_{c}^{o},\\
  \label{the heat in the cold isothermal process}
    Q_{c} =& -T_{c}\Delta S^{qs} - W_{c}^{irr} + \Delta K_{c}.
\end{align}
%using Eq.~(\ref{quasistatic and finite-time entropy changes in the cold isothermal process}).
Using Eqs.~(\ref{definition of work in isothermal process})
and (\ref{the heat in the cold isothermal process}),
we derive the output work in the cold isothermal process as
\begin{equation}
  \label{work in the cold isothermal process}
  W_{c} = -T_{c}\Delta S^{qs} - W_{c}^{irr} + \Delta K_{c} - \Delta E_{c},
\end{equation}
where $\Delta E_{c}$
%$=E(t_{cyc})-E(t_h)$ 
represents the internal energy change in this process.

As the cycle closes, the internal energy change per cycle in the particle vanishes.
From the first law of thermodynamics, we derive the output work per cycle as
\begin{equation}
  \label{output work per cycle}
  \begin{split}
    W %=& W^{iso}_{h} + W^{iso}_{c} + W^{ad}_{h\to c} + W^{ad}_{c\to h}, \\
    =& Q_{h} + Q_{c}\\
    =&\Delta T\Delta S^{qs} -W^{irr}_{h}-W^{irr}_{c},
  \end{split}
\end{equation}
using Eqs.~(\ref{work in hot isothermal process}), 
(\ref{kinetic energy change in the cold isothermal process}), and
(\ref{work in the cold isothermal process}).
%--(\ref{energy conservation of one cycle}).
%(\ref{work of adiabatic process from h to c}),
%(\ref{work of adiabatic process from c to h}),
As mentioned above, the irreversible works arise from the irreversibility 
of the isothermal processes.
If the irreversible works in Eq.~(\ref{output work per cycle}) vanish,
the work will be the same as $W^{qs}$ in Eq.~(\ref{quasistatic work with adiabatic condition}).
%The irreversible works arises due to the isothermal processes
%being finite in time and thus irreversible.
Thus, we call $W^{irr}_{h,c}$ the irreversible works as
the difference between $W$ in Eq.~(\ref{output work per cycle}) and $W^{qs}$. 
Using Eqs.~(\ref{the heat in the hot isothermal process})
and (\ref{output work per cycle}), we obtain the efficiency $\eta$ and power $P$ of the Carnot cycle as follows:
\begin{align}
  \label{efficiency}
  \eta \equiv \frac{W}{Q_{h}} =\frac{\Delta T \Delta S^{qs}
    -W^{irr}_{h}-W^{irr}_{c}}
       {T_{h}\Delta S^{qs} -W^{irr}_{h} + \Delta K_{h}},\\
  \label{power}
  P\equiv \frac{W}{t_{cyc}} = \frac{\Delta T\Delta S^{qs}
    -W^{irr}_{h}-W^{irr}_{c}}{t_{cyc}}.
\end{align}

\subsection{Small relaxation-times regime}
We consider the Carnot cycle in the regime where the relaxation times $\tau_{v}$ and $\tau_{x}(t)$ $(0\leq t\leq t_{cyc})$ are sufficiently small, which is of our main interest.
From Eq.~(\ref{sigma_x with small relaxation times}) in the Appendix,
the kinetic energy in this regime is approximated by
\begin{align}
  \frac{1}{2}m\sigma_{v}(0) =& \frac{1}{2}m\sigma_{v}(t_{cyc}) \simeq \frac{1}{2}T_{c},\\
  \frac{1}{2}m\sigma_{v}(t_{h}) \simeq& \frac{1}{2}T_{h}.
\end{align}
Thus, the kinetic energy change in the isothermal processes is given by
\begin{equation}
  \label{kinetic energy change in small relaxation times limit}
  \begin{split}
    \Delta K_{h}=-\Delta K_{c}
    \simeq& \frac{\Delta T}{2},
  \end{split}
\end{equation}
similarly to the quasistatic case, 
where we used Eq.~(\ref{kinetic energy change in the cold isothermal process}).
From Eqs.~(\ref{the heat in the hot isothermal process}), 
(\ref{the heat in the cold isothermal process}), and
(\ref{kinetic energy change in small relaxation times limit}),
the heat in the isothermal processes can be evaluated as follows:
\begin{equation}
  \label{supplied heat in the hot isothermal process 2}
  Q_{h} \simeq  T_{h}\Delta S^{qs} -W^{irr}_{h} + \frac{\Delta T}{2} ,
\end{equation}
\begin{equation}
  \label{supplied heat in the cold isothermal process 2}
  Q_{c} \simeq - T_{c}\Delta S^{qs} -W^{irr}_{c}-\frac{\Delta T}{2} .
\end{equation}
From Eq.~(\ref{efficiency}),
the efficiency in the small relaxation-times regime is given by
\begin{equation}
  \label{efficiency 2}
  \eta \simeq\frac{\Delta T \Delta S^{qs}
    -W^{irr}_{h}-W^{irr}_{c}}
       {T_{h}\Delta S^{qs} -W^{irr}_{h} + \frac{\Delta T}{2}}.
\end{equation}

Holubec and Ryabov pointed out the possibility of obtaining Carnot efficiency in a general class of finite-power Carnot cycle in the vanishing limit of the relaxation times~\cite{PhysRevE.96.062107,PhysRevLett.121.120601}.
In our underdamped Brownian Carnot cycle, we have to consider the heat leakage [$\Delta T/2$ in the denominator in Eq.~(\ref{efficiency 2})] because the kinetic energy cannot be neglected.
Thus, it may be impossible to achieve the Carnot efficiency in our finite-power Carnot cycle.
Nevertheless, if  $W_h^{\rm irr}$ and $W_c^{\rm irr}$ vanish in the vanishing limit of the relaxation times, the efficiency will reach the quasistatic efficiency in Eq.~(\ref{quasistatic efficiency under the adiabatic condition}), and we can achieve the Carnot efficiency as seen from Eq.~(\ref{efficiency in the small temperature difference}) in the small temperature-difference regime.
Subsequently, we study how the efficiency and power
depend on the relaxation times and temperature difference
in Sec.~\ref{numerical simulation}.

\section{Numerical simulations}
\label{numerical simulation}

In this section, we show the results of efficiency
and power obtained through the numerical
simulations of the proposed Brownian Carnot cycle
as varying the relaxation times and temperature difference.
In these simulations, we solved
Eqs.~(\ref{equation of sigma_x})--(\ref{equation of sigma_xv})
numerically by using the fourth-order Runge-Kutta method.
The specific protocol $\lambda(t)$ for our simulations
is given by
\begin{equation}
  \label{optimized protocol in the overdamped regime}
  \lambda(t)=\left\{
  \begin{split}
    &\frac{T_h}{\sigma_a(1+b_{1}\frac{t}{t_{h}})^2}
    \hspace{3mm} (0\leq t\leq  t_{h})\\
    &\frac{T_c}{\sigma_b(1+b_{2}\frac{t-t_{h}}{t_{c}})^2}
    \hspace{2mm}(t_{h}\leq  t\leq  t_{cyc}),
  \end{split}
  \right.
\end{equation}
where $\sigma_{a}$ and $\sigma_{b}$ ($>\sigma_{a}$) are positive constants,
and we defined $b_{1} \equiv \sqrt{\sigma_{b}/\sigma_{a}}-1$ and
$b_{2}\equiv \sqrt{\sigma_{a}/\sigma_{b}}-1$.
This protocol is inspired by the optimal protocol
in the overdamped Brownian Carnot cycle~\cite{Schmiedl_2007,PhysRevE.96.062107}
and satisfies Eq.~(\ref{condition of the adiabatic process to achieve the Carnot efficiency}) assigned to the protocol.
This protocol also satisfies
the scaling condition in Eq.~(\ref{assumption of the protocol}).
For all the simulations,
we fixed $\sigma_{b}/\sigma_{a} = 2.0$, $T_{c} = 1.0$, $t_{h}=t_{c}=1.0$,
and $\gamma = 1.0$ 
and varied the temperature difference $\Delta T$, or equivalently, the temperature $T_{h}$.
We calculated the heat in Eqs.~(\ref{supplied heat in the hot isothermal process}) and (\ref{supplied heat in the cold isothermal process}) and the work $W=Q_h+Q_c$ in Eq.~(\ref{output work per cycle}) from the solution of Eqs.~(\ref{equation of sigma_x})--(\ref{equation of sigma_xv}).
Using the heat and work,
we also numerically calculated the efficiency $\eta=W/Q_h$ using Eq.~(\ref{efficiency})
and power $P=W/t_{cyc}$ using Eq.~(\ref{power}).
Before starting to measure the thermodynamic quantities,
we waited until the system settled down to a steady cycle. Moreover, when we take the limit $m\to 0$,
the relaxation time of velocity $\tau_{v}=m/\gamma$ vanishes.
By a simple calculation from
Eqs.~(\ref{relaxation time of x}) and (\ref{optimized protocol in the overdamped regime}), we find that $\tau_{x}$ satisfies
\begin{equation}
  \frac{\gamma\sigma_{a}}{T_{h}}\leq \tau_{x}(t)\leq \frac{\gamma\sigma_{b}}{T_{c}}.
\end{equation}
Thus, the smaller $\sigma_{a}$ and $\sigma_{b}$ are,
the smaller $\tau_{x}$ is.
When we take the limit $\sigma_{a},\sigma_{b} \to 0$
while maintaining $\sigma_{b}/\sigma_{a}$ finite, $\tau_{x}(t)$ vanishes and $\lambda(t)=\gamma/\tau_x(t)$ from Eq.~(\ref{relaxation time of x}) diverges.
Because $\tau_x(0)\propto \sigma_a$ and $\tau_v\propto m$ are satisfied, we varied the mass $m$ and the parameter $\sigma_{a}$ to vary the relaxation times.
Note that in the numerical simulations, we selected a time step smaller than the relaxation times. 
Specifically, we set the time step as $\min(m,\sigma_a)\times 10^{-2}$ because of $\tau_x(0)\propto \sigma_a$ and $\tau_v\propto m$.

To evaluate the efficiency in Eq.~(\ref{efficiency}) obtained numerically,
we compared it with the quasistatic efficiency $\eta^{qs}$ in Eq.~(\ref{quasistatic efficiency under the adiabatic condition}).
Because $\eta_C$ in Eq.~(\ref{Carnot efficiency}) is proportional to $\Delta T$, the ratio of $\eta^{qs}$ in Eq.~(\ref{efficiency in the small temperature difference}) to $\eta_C$ in the small temperature-difference regime satisfies
\begin{equation}
\label{ratio of the quasistatic efficiency to the Carnot one}
    \frac{\eta^{qs}}{\eta_C}=1-O(\Delta T).
\end{equation}

\begin{figure}[t]
  \includegraphics[width=6.2cm]{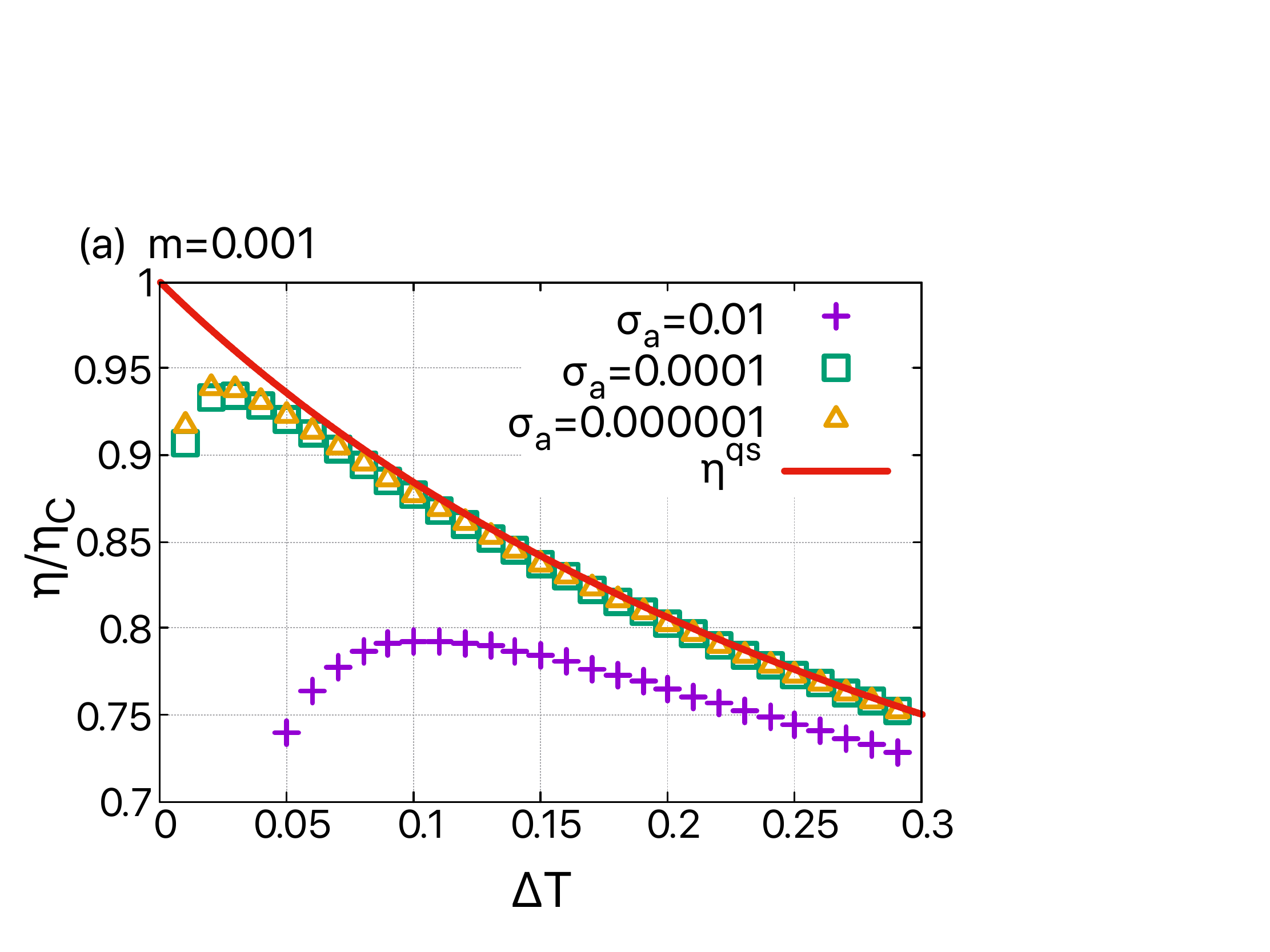}
  \includegraphics[width=6.2cm]{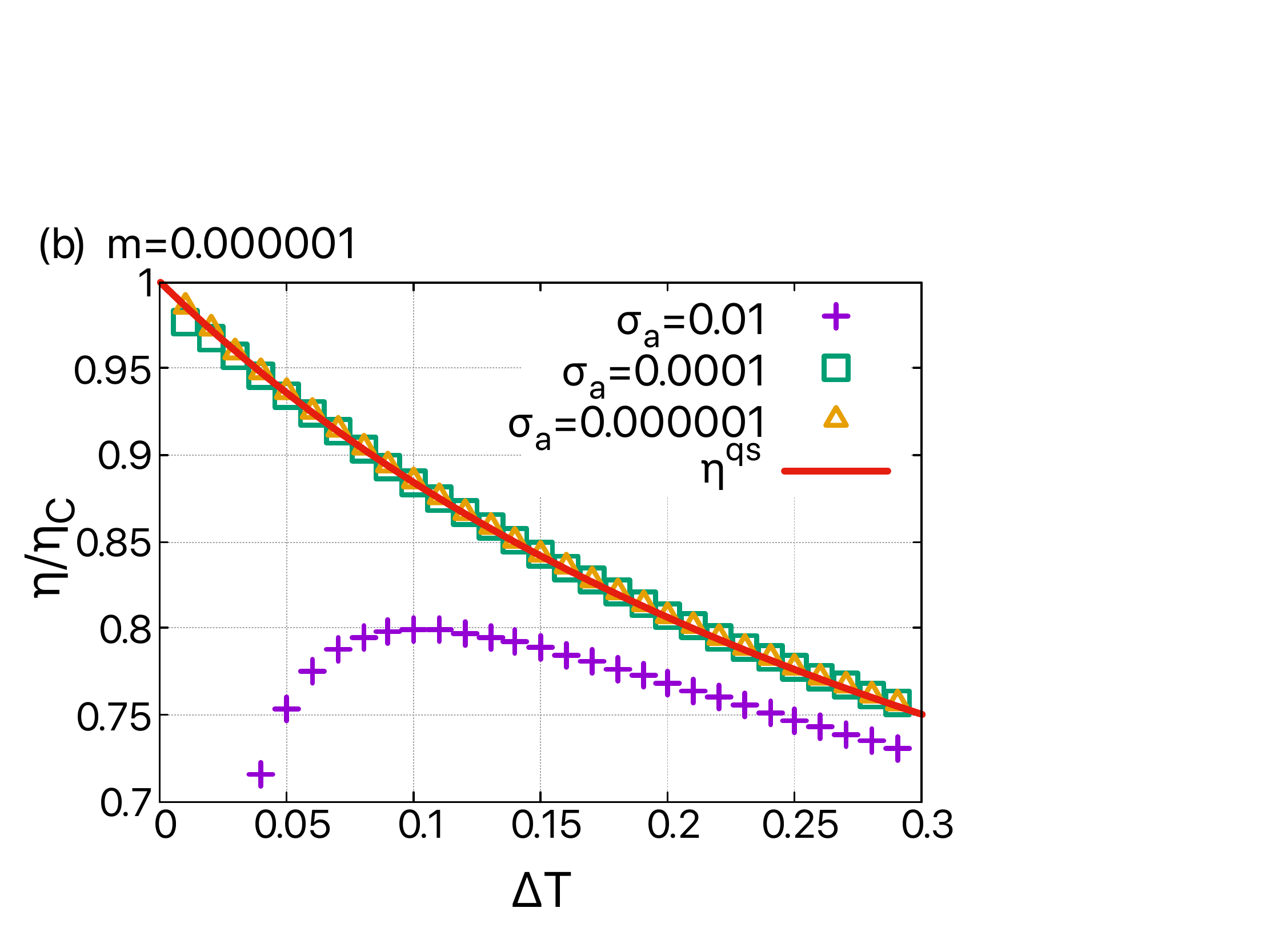}
  \caption{
    The ratio of the efficiency in Eq.~(\ref{efficiency}) to the
    Carnot efficiency in our cycle
    with the protocol in Eq.~(\ref{optimized protocol in the overdamped regime}) when $\tau_{x}$ varies
    at (a) $\tau_{v}=10^{-3}$ and (b) $\tau_{v}=10^{-6}$.
    Because the parameter $\sigma_{a}$ is proportional to $\tau_{x}(0)$ in the protocol in Eq.~(\ref{optimized protocol in the overdamped regime}), we vary $\sigma_{a}$ to make $\tau_x$ small.
    Similarly, we vary the mass $m$ because it is proportional to $\tau_v$.
    In these simulations, we set $\sigma_{a}=10^{-2}$ (purple plus),
    $\sigma_{a}=10^{-4}$ (green square),
    and $\sigma_{a}=10^{-6}$ (orange triangle).
    The red solid line corresponds to the ratio of $\eta^{qs}$
    in Eq.~(\ref{quasistatic efficiency under the adiabatic condition})
    to the Carnot efficiency.
    The efficiency appears to approach the Carnot efficiency in the vanishing limit
    of $\sigma_{a}$ (or $\tau_{x}$), $m$ (or $\tau_{v}$), and $\Delta T$.
  }
  \label{fig:efficiency}
\end{figure}
\begin{figure}[t]
  \vspace{-2mm}
  \includegraphics[width=6.3cm]{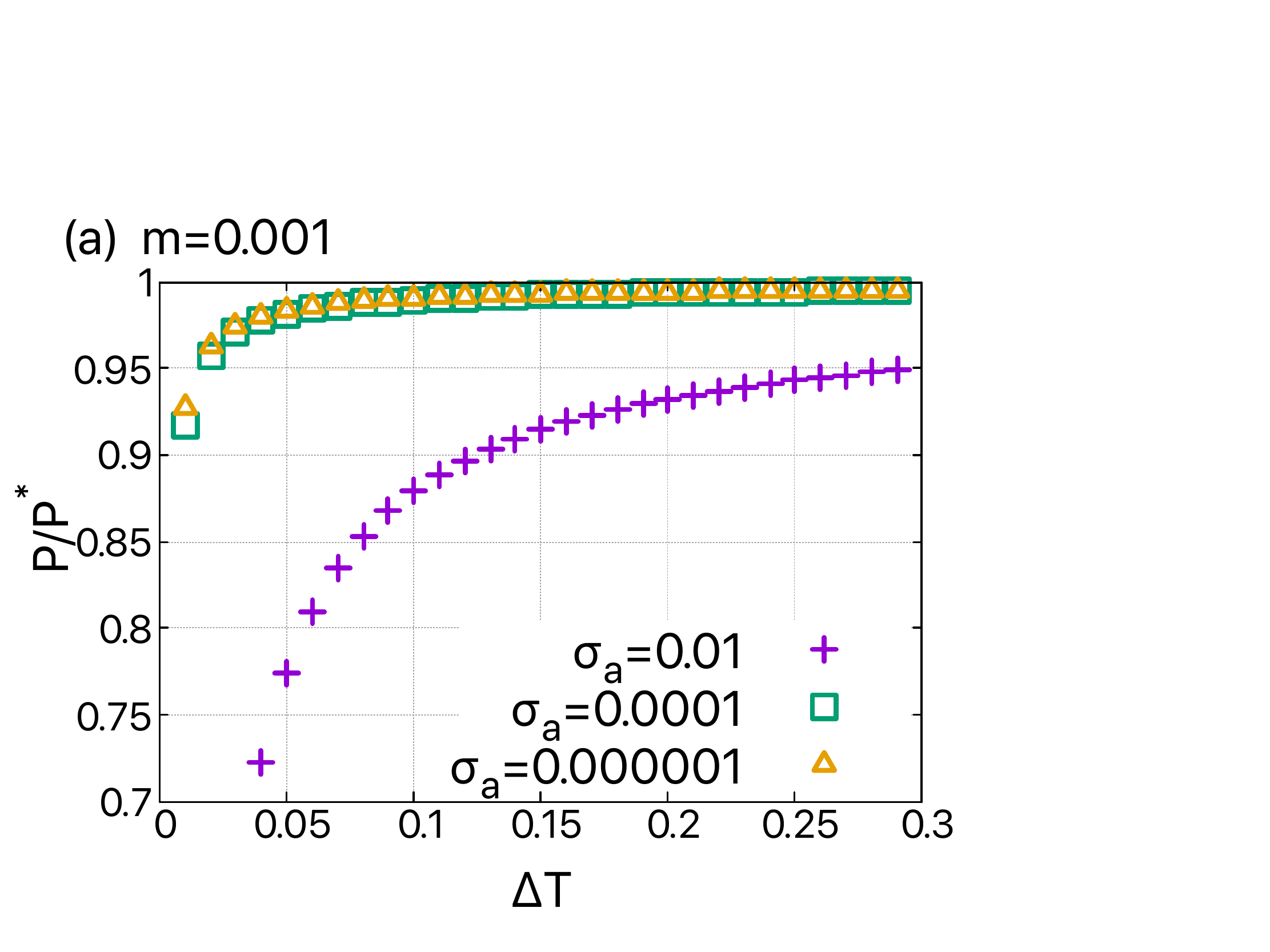}
  \includegraphics[width=6.3cm]{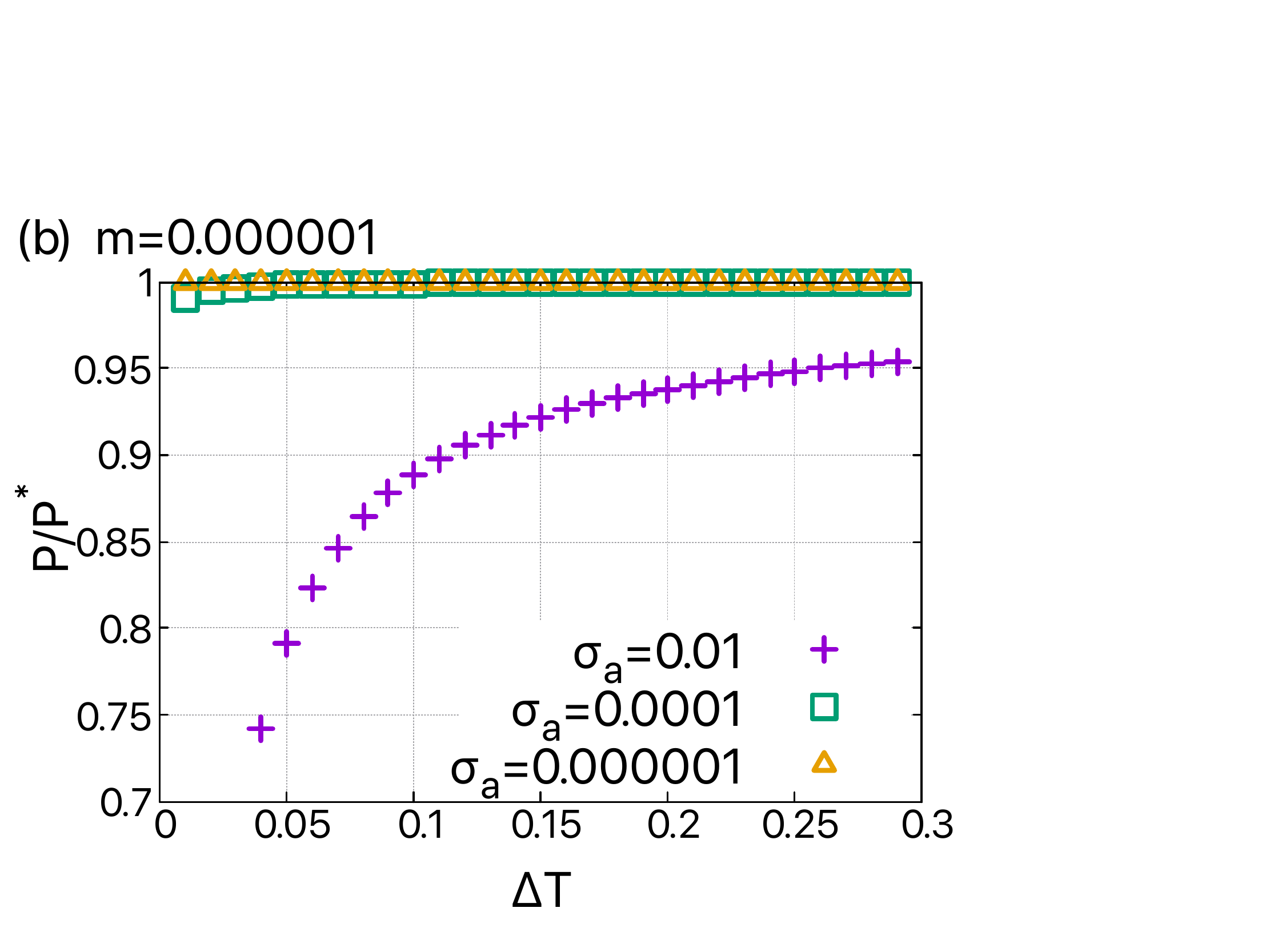}
  \caption{
    The ratio of the power in Eq.~(\ref{power})
    to $P^{*}$ in Eq.~(\ref{quasistatic power}) in the proposed cycle
    corresponding to Figs.~\ref{fig:efficiency}(a) and \ref{fig:efficiency}(b).
    %Since the parameter $\sigma_{a}$ is proportional to $\tau_{x}(0)$ in the protocol Eq.~(\ref{optimized protocol in the overdamped regime}), we vary $\sigma_{a}$ to make $\tau_x$ be small.
    %Similarly, we vary the mass $m$ because it is proportional to $\tau_v$.
    The power appears to approach $P^{*}$ in Eq.~(\ref{quasistatic power})
    in the vanishing limit
    of $\sigma_{a}$ (or $\tau_{x}$),  $m$ (or $\tau_{v}$), and $\Delta T$.
  }
  \label{fig:power}
\end{figure}

Similarly, we evaluate the power in Eq.~(\ref{power})
by using a criterion $P^{*}$ defined as follows:
\begin{equation}
  \label{quasistatic power}
  P^{*} \equiv \frac{W^{qs}}{t_{cyc}}
  = \frac{\Delta T\Delta S^{qs}}{t_{cyc}},
\end{equation}
where $W^{qs}$ is the quasistatic work in Eq.~(\ref{quasistatic work}).
Here, we regard the power as finite
when the power in Eq.~(\ref{power}) 
is the same order as $P^{*}$.

Figure \ref{fig:efficiency} shows the ratio of the efficiency of the proposed cycle with the protocol in Eq.~(\ref{optimized protocol in the overdamped regime}) to the Carnot efficiency.
We can see that the efficiency approaches $\eta^{qs}$
with $\tau_{x},\ \tau_{v} \to 0$.
Considering Eqs.~(\ref{quasistatic efficiency under the adiabatic condition}) and (\ref{efficiency 2}), we can expect that the irreversible works disappear. Thus, the efficiency can be regarded as the Carnot efficiency in the small relaxation-times and small temperature-difference regime.

Figure \ref{fig:power} shows the ratio of
the power to $P^{*}$ in Eq.~(\ref{quasistatic power}),
corresponding to Fig.~\ref{fig:efficiency}.
At any $\Delta T$, we can see that the power approaches $P^{*}$
as $\tau_{x},\ \tau_{v}\to 0$ .
As the power in Eq.~(\ref{power}) is defined using the work in Eq.~(\ref{output work per cycle}), the ratio of $P$ to $P^{*}$ is the same as the ratio of $W$ to $W^{qs}$ in Eq.~(\ref{quasistatic work}).
When the power  $P$ approaches $P^{*}$, the work $W$ approaches $W^{qs}$.
This implies that the irreversible works vanish.
Because the power is of the same order as $P^{*}$ from Fig.~\ref{fig:power}, we can consider the power to be finite.
Therefore, Figs.~\ref{fig:efficiency} and \ref{fig:power} imply
that the Carnot efficiency and finite power are compatible
in the vanishing limit of the relaxation times
in the small temperature-difference regime.

\section{Theoretical analysis}
\label{theoretical analysis}
This section analytically shows that it is possible to achieve the Carnot efficiency in our cycle in the vanishing limit of the relaxation times in the small temperature-difference regime without breaking the trade-off relation in Eq.~(\ref{general trade-off relation}), as implied in the numerical results in Sec. \ref{numerical simulation}.

In general, the efficiency decreases when the entropy production increases, as shown in Eq.~(\ref{efficiency with entropy production}).
As the adiabatic processes have no entropy production because no heat exchange is present, we have only to consider the entropy production in the isothermal processes.
In the small relaxation-times regime, the efficiency in Eq.~(\ref{efficiency}) is approximated by that in Eq.~(\ref{efficiency 2}).
If $W^{irr}_{h,c}\to 0$ is satisfied in the vanishing limit of the relaxation times, the efficiency in Eq.~(\ref{efficiency 2}) approaches the quasistatic efficiency in Eq.~(\ref{quasistatic efficiency under the adiabatic condition}).
As seen in Eq.~(\ref{efficiency in the small temperature difference}), it is expected that the contribution of the heat leakage to the efficiency can be neglected in the small temperature-difference regime.
Thus, the efficiency in Eq.~(\ref{efficiency}) approaches the Carnot efficiency in the small relaxation-times and small temperature-difference regime, and the power in Eq.~(\ref{power}) also approaches $P^{*}$ in Eq.~(\ref{quasistatic power}) simultaneously.

The numerical results imply that the irreversible works vanish
in the vanishing limit of the relaxation times $\tau_{x}$ and $\tau_{v}$.
To derive a similar conclusion analytically, we first show that the irreversible works relate to the entropy production given by
\begin{equation}
  \label{definition of the entropy production}
  \Sigma \equiv -\frac{Q_{h}}{T_{h}} -\frac{Q_{c}}{T_{c}}.
\end{equation}
Similarly to $\Sigma^{qs}$ in Eq.~(\ref{entropy production in quasistatic cycle}),
the entropy production $\Sigma$ is expressed only by an
entropy change in the heat baths.
In the small relaxation-times regime,
we can express $\Sigma$ in Eq.~(\ref{definition of the entropy production}) as follows:
\begin{align}
  \label{the entropy production with the irreversible works}
    \Sigma \simeq & \frac{-\frac{\Delta T}{2} - T_{h}\Delta S^{qs} +W^{irr}_{h}}{T_{h}}
    +\frac{\frac{\Delta T}{2} + T_{c}\Delta S^{qs}  + W^{irr}_{c}}{T_{c}}\nonumber\\
    =&\frac{W^{irr}_{h}}{T_{h}}+\frac{W^{irr}_{c}}{T_{c}} +\frac{(\Delta T)^2}{2T_{h}T_{c}},
\end{align}
using Eqs.~(\ref{supplied heat in the hot isothermal process 2})
and (\ref{supplied heat in the cold isothermal process 2}).
The last term on the right-hand side
of Eq.~(\ref{the entropy production with the irreversible works})
comes from the heat leakage due to the instantaneous adiabatic processes.
From Eq.~(\ref{the entropy production with the irreversible works}),
the entropy production can be regarded as zero in the small temperature-difference regime
when the irreversible works vanish.
In general, the entropy production
in Eq.~(\ref{definition of the entropy production}) can also be rewritten as
\begin{equation}
  \label{the entropy production and the efficiency 1}
  \begin{split}
    \Sigma %=& -\frac{Q_{h}}{T_{h}}-\frac{W-Q_{h}}{T_{c}}\\
    %=& Q_{h}\left\{-\frac{1}{T_{h}} -\frac{1}{T_{c}}\frac{W}{Q_{h}} +\frac{1}{T_{c}}\right\}\\
    =&\frac{Q_{h}}{T_{c}}(\eta_{C} - \eta),
  \end{split}
\end{equation}
where we used Eqs.~(\ref{Carnot efficiency}) and (\ref{efficiency}).
This equation shows that the efficiency approaches the Carnot efficiency
when the entropy production vanishes.
Thus, by using Eqs.~(\ref{the entropy production with the irreversible works})
and (\ref{the entropy production and the efficiency 1}),
we obtain the efficiency as
\begin{align}
  \label{efficiency with entropy production}
    \eta &= \eta_{C}- \frac{T_{c}\Sigma}{Q_{h}}\nonumber\\
    &\simeq\eta_{C}- \frac{T_{c}}{Q_{h}}\left(\frac{W^{irr}_{h}}{T_{h}}+\frac{W^{irr}_{c}}{T_{c}}\right)
    + O[(\Delta T)^2],
\end{align}
in the small relaxation-times regime.
Here, the contribution of the heat leakage
to the efficiency is $O[(\Delta T)^2]$, and it is negligible in the small temperature-difference regime.

We consider the trade-off relation in Eq.~(\ref{general trade-off relation}) to discuss the compatibility of the Carnot efficiency and finite power in our Brownian Carnot cycle.
Using Eq.~(\ref{the entropy production and the efficiency 1}),
we can rewrite Eq.~(\ref{general trade-off relation}) as
\begin{equation}
\label{trade-off relation in our cycle}
  P\leq \frac{\eta T_{c}}{Q_{h}}A\Sigma
\end{equation}
in terms of the entropy production $\Sigma$.
When the quantity $A\Sigma$ is nonzero in the vanishing limit of the entropy production $\Sigma$, implying that $A$ should diverge, the finite power may be allowed.
In fact, when the entropy production $\Sigma$ vanishes in the small temperature-difference regime, the irreversible works should vanish because of Eq.~(\ref{the entropy production with the irreversible works}).
Then, the power in Eq.~(\ref{power}) approaches $P^{*}$ in Eq.~(\ref{quasistatic power}), which implies that the power is regarded as finite.
Thus, we find the expression $A\Sigma$ in our cycle below.

\subsection{Trade-off relation between power and efficiency}
We derive the trade-off relation in our cycle.
To obtain the expression of the entropy production,
we use Eqs. (67) and (68)
from Ref.~\cite{PhysRevE.97.062101}.
Using the general expression of the entropy production
in the Langevin system~\cite{Seifert_2012,PhysRevE.85.051113},
Dechant and Sasa showed a trade-off relation for the underdamped Langevin system in Ref.~\cite{PhysRevE.97.062101}.
Thus, we can apply their results to our system.
Applying Eq. (67) from Ref.~\cite{PhysRevE.97.062101},
we can divide the probability currents in Eqs.~(\ref{the probability current of x})
and (\ref{the probability current of v})
into the reversible parts, $j_{x}^{rev}$ and $j_{v}^{rev}$,
and the irreversible parts, $j_{x}^{irr}$ and $j_{v}^{irr}$, as
\begin{equation}
  \label{probability currents of x and v}
  \begin{split}
    j_{x}(x,v,t)=&j_{x}^{rev}(x,v,t)+j_{x}^{irr}(x,v,t),\\
    j_{v}(x,v,t)=&j_{v}^{rev}(x,v,t)+j_{v}^{irr}(x,v,t),
  \end{split}
\end{equation}
where
\begin{equation}
  \begin{split}
    \label{divided probability currents}
    j^{rev}_{x}(x,v,t)\equiv&v p(x,v,t),\hspace{1cm} j^{irr}_{x}(x,v,t)\equiv 0,\\
    j^{rev}_{v}(x,v,t) \equiv &-\frac{\lambda(t)}{m}x p(x,v,t),\\
    j^{irr}_{v}(x,v,t) \equiv &\left(-\frac{\gamma}{m}v
    - \frac{\gamma T(t)}{m^2}\frac{\partial}{\partial v}\right)p(x,v,t).
  \end{split}
\end{equation}
%Here, $T$ depends on time as
%\begin{equation}
%T(t)=\left\{
%\begin{split}
%&T_{h} \hspace{3mm} (0 < t< t_{h})\\
%&T_{c} \hspace{3mm} (t_{h} < t< t_{cyc}).
%\end{split}
%\right.
%\end{equation}
For convenience, we introduce a function $\phi(t)$ to describe
the time evolution of the temperature as
\begin{equation}
  \label{time dependence of the temperature}
  \begin{split}
    \frac{1}{T(t)}=&\frac{1}{T_{c}}-
    \left(\frac{1}{T_{c}} - \frac{1}{T_{h}}\right)\phi(t)\\
    =&\frac{1}{T_{c}}\left[1-\eta_{C}\phi(t)\right].
  \end{split}
\end{equation}
In our cycle, the function $\phi(t)$ is given by
\begin{equation}
  \label{time dependence of temperature phi}
  \phi(t)\equiv\left\{
  \begin{split}
    &1\hspace{3mm} (0 < t < t_{h})\\
    &0\hspace{3mm} (t_{h} < t < t_{cyc}).
  \end{split}
  \right.
\end{equation}
Using Eq.~(\ref{Fokker-Planck equation}), the heat flux in Eq.~(\ref{heat flux expressed by variables}) is rewritten as
\begin{align}
  \label{heat flux with the probability currents}
      \dot{Q}=&\int^{\infty}_{-\infty} dx\int^{\infty}_{-\infty} dv \left[\frac{1}{2}mv^2 + \frac{1}{2}\lambda x^2 \right]\frac{\partial p}{\partial t}\nonumber\\
    =& -\int^{\infty}_{-\infty} dx\int^{\infty}_{-\infty} dv \left[\frac{1}{2}mv^2 + \frac{1}{2}\lambda x^2 \right]
    \left(\frac{\partial j_{x}}{\partial x} + \frac{\partial j_{v}}{\partial v}\right)\nonumber\\
    =& \int^{\infty}_{-\infty} dx\int^{\infty}_{-\infty} dv \left[m v j_{v} + \lambda x j_{x}\right],
\end{align}
where the last equality is derived from the integration by parts and we 
assumed that the probability currents at the boundary vanish.
By using Eqs.~(\ref{probability currents of x and v}), 
(\ref{divided probability currents}), and (\ref{heat flux with the probability currents}),
we obtain the heat flux as
\begin{equation}
  \label{the heat flux with irreversible current}
  \begin{split}
    \dot{Q}(t) =& \int^{\infty}_{-\infty} dx\int^{\infty}_{-\infty} dv\ mv j_{v}^{irr}(x,v,t).
  \end{split}
\end{equation}
Thus, we obtain the heat flowing from the heat bath to
the Brownian particle in the hot isothermal process as
\begin{equation}
  \label{the heat with irreversible current}
  \begin{split}
    Q_{h} =& \int^{t_{h}}_{0}dt\ \int^{\infty}_{-\infty} dx\int^{\infty}_{-\infty} dv\
    mv j_{v}^{irr}\\
    =&\int^{t_{cyc}}_{0}dt\ \int^{\infty}_{-\infty} dx\int^{\infty}_{-\infty} dv\
    \phi(t)mv j_{v}^{irr},
  \end{split}
\end{equation}
using Eq.~(\ref{time dependence of temperature phi}).
Now, we consider the entropy production rate.
Based on Eq. (68) from Ref.~\cite{PhysRevE.97.062101},
the entropy production rate is given by \cite{Seifert_2012,PhysRevE.85.051113}
\begin{equation}
  \label{definition of the entropy production rate}
  \begin{split}
    \dot{\Sigma}(t) =& \int^{\infty}_{-\infty} dx\int^{\infty}_{-\infty} dv\
    \frac{m^{2}(j_{v}^{irr}(x,v,t))^2}{\gamma T(t) p(x,v,t)}.
  \end{split}
\end{equation}
Using Eq.~(\ref{definition of the entropy production rate}),
we can also obtain the concrete expression of the entropy production per cycle as
\begin{equation}
  \label{definition of entropy production per cycle}
  \begin{split}
    \Sigma =&\int^{t_{cyc}}_{0}dt\ \dot{\Sigma}(t)\\
    =&\int^{t_{cyc}}_{0}dt\ \int^{\infty}_{-\infty} dx\int^{\infty}_{-\infty} dv\
    \frac{m^{2}(j_{v}^{irr}(x,v,t))^2}{\gamma T(t) p(x,v,t)}.    
  \end{split}
\end{equation}
From the Cauchy–Schwarz inequality, it is shown that the upper bound of the heat flux in Eq.~(\ref{the heat flux with irreversible current}) is expressed using the entropy production rate as
\begin{equation}
  \label{entropic bound of the heat flux}
  \begin{split}
    \dot{Q}^2 =& \left(\int^{\infty}_{-\infty} dx\int^{\infty}_{-\infty} dv\ v\sqrt{\gamma Tp}\frac{m j_{v}^{irr}}{\sqrt{\gamma Tp}}\right)^2\\
    \leq& \left(\int^{\infty}_{-\infty} dx\int^{\infty}_{-\infty} dv\ \gamma T v^2p \right)\\
    &\times\left(\int^{\infty}_{-\infty} dx\int^{\infty}_{-\infty} dv\ \frac{m^2(j_{v}^{irr})^2}{\gamma T p}\right)\\
    =&\gamma T\sigma_{v}\dot{\Sigma},
  \end{split}
\end{equation}
or, equivalently,
\begin{equation}
  \label{entropic bound of the absolute value of the heat flux}
  |\dot{Q}|\leq \sqrt{\gamma T\sigma_{v}\dot{\Sigma}}.
\end{equation}
Because $\gamma T(t)\sigma_{v}$ and $\dot{\Sigma}$ are positive, by using Eq.~(\ref{entropic bound of the absolute value of the heat flux}), we can derive the following bound for the heat in Eq.~(\ref{the heat with irreversible current}):
\begin{equation}
  \label{entropic bound of the heat}
  \begin{split}
    (Q_{h})^2 =& \left(\int^{t_{cyc}}_{0}dt\ \phi(t)\
    \dot{Q}(t)\right)^2\\
    \leq&\left(\int^{t_{cyc}}_{0}dt\ 
    \phi(t)\sqrt{\gamma T(t)\sigma_{v}\dot{\Sigma}}\right)^2\\
    \leq&\left(\int^{t_{cyc}}_{0}dt\
    \phi^2(t)\gamma T(t)\sigma_{v}\right)
    \left(\int^{t_{cyc}}_{0}dt\ \dot{\Sigma}\right)\\
    =&t_{cyc}T_{c}^2\chi \Sigma,
  \end{split}
\end{equation}
where
\begin{equation}
  \label{amplitude}
  \chi \equiv \frac{\gamma}{t_{cyc}T_{c}}\int^{t_{cyc}}_{0}dt\
  \frac{\phi^2(t)}{1-\eta_{C}\phi(t)}\sigma_{v}(t),
\end{equation}
and we used the Cauchy–Schwarz inequality and Eq.~(\ref{time dependence of the temperature}).
Using Eqs.~(\ref{the entropy production and the efficiency 1}) and (\ref{entropic bound of the heat}),
we can derive the trade-off relation 
in our cycle as
\begin{equation}
  \label{trade-off relation in the Brownian Carnot cycle}
  \begin{split}
    P =&\frac{W}{t_{cyc}}= \frac{W}{Q_{h}}\frac{1}{Q_{h}}\frac{Q_{h}^2}{t_{cyc}}\\
    \leq& \eta\frac{1}{Q_{h}}T_{c}^2\chi\Sigma\\
    =&\chi T_{c}\eta(\eta_{C}-\eta).
  \end{split}
\end{equation}
By comparing Eqs.~(\ref{trade-off relation in our cycle}) and (\ref{trade-off relation in the Brownian Carnot cycle}), we obtain $A=T_c \chi$.
We will show that in the limit of $\tau_{x},\tau_{v}\to 0$,
the entropy production $\Sigma$ vanishes and $\chi$ diverges
while $\chi \Sigma$ maintains positive.
For this purpose,
we rewrite Eq.~(\ref{definition of the entropy production rate}) as follows.
In our model (Sec.~\ref{model}),
the probability distribution was assumed to be
%the form of Eq.~(\ref{Gaussian distribution}).
the Gaussian distribution shown in Eq.~(\ref{Gaussian distribution}).
Thus, we can differentiate the distribution function $p(x,v,t)$ with respect to $v$ as
\begin{equation}
  \label{partial derivative with respect to v}
  \frac{\partial p}{\partial v}=\frac{\sigma_{xv}x-\sigma_{x}v}{\sigma_{x}\sigma_{v}-\sigma_{xv}^2}p.
\end{equation}
We can rewrite the entropy production rate in Eq.~(\ref{definition of the entropy production rate}) by using the variables $\sigma_{x}$, $\sigma_{v}$, and $\sigma_{xv}$
and derive the expression of $\dot\Sigma$ under the assumption of the Gaussian distribution as
\begin{align}
  \label{rewrite entropy production rate 1}
    \dot{\Sigma}(t)
    =&\int^{\infty}_{-\infty} dx \int^{\infty}_{-\infty}dv\ \frac{m^{2}}{\gamma T p}
    \left\{\left(\frac{\gamma}{m} v + \frac{\gamma T}{m^2}
    \frac{\partial}{\partial v}\right)p\right\}^2\nonumber\\
    =&\frac{m^2}{\gamma T}\int^{\infty}_{-\infty} dx\int^{\infty}_{-\infty} dv
    \left\{\frac{\gamma}{m}v+\frac{\gamma T}{m^2}
    \frac{\sigma_{xv}x-\sigma_{x}v}{\sigma_{x}\sigma_{v}-\sigma_{xv}^2}\right\}^2p\nonumber\\
    %=& \frac{m^2}{\gamma T}
    %\left[ \frac{\gamma^2}{m^2}\sigma_{v}
    %- 2\frac{\gamma^2 T}{m^3}
    %+\frac{\gamma^2 T^2}{m^4}
    %\frac{\sigma_x}{\sigma_{x}\sigma_{v} - \sigma_{xv}^2}\right]\\
    =&\frac{\frac{\gamma}{m}\left(T-m\sigma_v\right)^2
      + (2T-m\sigma_{v})\gamma\frac{\sigma_{xv}^2}{\sigma_{x}}}
    {T\left(m\sigma_{v}-\tau_{v}\gamma
      \frac{\sigma_{xv}^2}{\sigma_{x}}\right)},
\end{align}
where we used Eqs.~(\ref{relaxation time of x}),
(\ref{relaxation time of v}),
(\ref{equation of sigma_v}),
(\ref{heat flux expressed by variables}),
(\ref{divided probability currents}), 
and (\ref{partial derivative with respect to v}).
Using Eqs.~(\ref{equation of sigma_v}) and (\ref{heat flux expressed by variables}),
we obtain 
\begin{equation}
    \dot{Q}=\frac{\gamma}{m}(T-m\sigma_v).
\end{equation}
Thus, Eq.~(\ref{rewrite entropy production rate 1}) can be rewritten as
\begin{equation}
  \label{rewrite entropy production rate 1-2}
  \begin{split}
    \dot{\Sigma}(t)=&\frac{\tau_{v}\dot{Q}^2
      + (2T-m\sigma_{v})\gamma\frac{\sigma_{xv}^2}{\sigma_{x}}}
    {T\left(m\sigma_{v}-\tau_{v}\gamma
      \frac{\sigma_{xv}^2}{\sigma_{x}}\right)}.
  \end{split}
\end{equation}
Integrating Eq.~(\ref{rewrite entropy production rate 1-2}) with respect to time,
we derive the entropy production per cycle $\Sigma$ in our cycle as
\begin{equation}
  \label{entropy production in our Carnot cycle}
  \Sigma=\int^{t_{cyc}}_{0}dt\frac{\tau_{v}\dot{Q}^2(t)
    + [2T(t)-m\sigma_{v}(t)]\gamma\frac{\sigma_{xv}^2(t)}{\sigma_{x}(t)}}
             {T(t)\left(m\sigma_{v}(t)-\tau_{v}\gamma
               \frac{\sigma_{xv}^2(t)}{\sigma_{x}(t)}\right)}.
\end{equation}

\subsection{Small relaxation-times regime}
We evaluate the entropy production in Eq.~(\ref{entropy production in our Carnot cycle}) in the small relaxation-times regime.
In the hot isothermal process, the process can be divided into the relaxation part and the part after the relaxation.
Because the relaxation time of the system at the beginning of the hot isothermal process is given by $\tau_{0}\equiv \max(\tau_{x}(0),\tau_{v})$, the entropy production in the hot isothermal process $\Sigma_h$ is divided as
\begin{equation}
\label{entropy production in the hot isothermal process}
    \Sigma_h 
    \equiv  \int^{t_{h}}_{0}dt\ \dot{\Sigma} 
    = \int^{\tau_0}_{0}dt\ \dot{\Sigma}+\int^{t_{h}}_{\tau_0}dt\ \dot{\Sigma},
\end{equation}
where the first and second terms in Eq.~(\ref{entropy production in the hot isothermal process}) represent the entropy production in the relaxation and after the relaxation, respectively.
%Since the relaxation is instantaneous in the vanishing limit of the relaxation times, the entropy production rate $\dot{\Sigma}$ in the relaxation should diverge, and 
We first evaluate the entropy production after the relaxation.
From Eqs.~(\ref{sigma_x with small relaxation times}) and (\ref{sigma_xv with small relaxation times}) in 
the Appendix, the variables $\sigma_x$, $\sigma_v$, and $\sigma_{xv}$ after the relaxation satisfy
\begin{equation}
  \label{approximation of variables}
  \sigma_{x} \simeq \frac{T}{\lambda},
  \quad
  \sigma_{v} \simeq \frac{T}{m},
  \quad
  \sigma_{xv} \simeq -\frac{T}{2\lambda^2}\frac{d\lambda}{dt}.
\end{equation}
%where we assumed that the relaxation times are much smaller than $t_{cyc}$.
Then, we can obtain
\begin{equation}
  \label{the heat flux with relaxation time of x}
  \gamma \frac{\sigma_{xv}^2(t)}{\sigma_{x}(t)}
  \simeq \frac{\tau_{x}(t)T}{4}\left(\frac{d}{dt}\ln \lambda(t)\right)^2.
  %=\frac{\tau_{x}(s)T}{4t_{cyc}^2}\left(\frac{d}{ds}\ln \Lambda(s)\right)^2.
\end{equation}
Using Eqs.~(\ref{rewrite entropy production rate 1-2}), (\ref{approximation of variables}), and (\ref{the heat flux with relaxation time of x}), the entropy production rate after the relaxation is given by
\begin{equation}
  \label{rewrite entropy production rate 2}
  \begin{split}
    \dot{\Sigma}(t)
    %\simeq&
    %\frac{1}{T}\frac{\tau_{v}\dot{Q}(t)^2
      %+\tau_{x}\frac{T^2}{4}\left(\frac{d}{dt}\ln \lambda\right)^2}
         %{T-\tau_{v}\tau_{x}
           %\frac{T}{4}\left(\frac{d}{dt}\ln \lambda\right)^2}\\
         \simeq&\frac{1}{t_{cyc}T}\frac{\frac{\tau_{v}}{t_{cyc}}\left(\frac{dQ(s)}{ds}\right)^2
           +\frac{\tau_{x}}{t_{cyc}}\frac{T^2}{4}\left(\frac{d}{ds}\ln \Lambda\right)^2}
         {T-\frac{\tau_{v}}{t_{cyc}}\frac{\tau_{x}}{t_{cyc}}
           \frac{T}{4}\left(\frac{d}{ds}\ln \Lambda\right)^2},
  \end{split}
\end{equation}
where we used $s=t/t_{cyc}$ to compare
the cycle time $t_{cyc}$ and the relaxation times $\tau_{x}$ and $\tau_{v}$.
Then, we derive the entropy production after the relaxation in the hot isothermal process as
\begin{equation}
\label{entropy production hot after the relaxation}
   \int^{t_{h}}_{\tau_0}dt\ \dot{\Sigma}
   \simeq \frac{1}{T_{h}}\int^{t_{h}/t_{cyc}}_{\tau_{0}/t_{cyc}}ds\
    \frac{\frac{\tau_{v}}{t_{cyc}}\left(\frac{dQ(s)}{ds}\right)^2
      +\frac{\tau_{x}}{t_{cyc}}\frac{T_{h}^2}{4}\left(\frac{d}{ds}\ln \Lambda\right)^2}
         {T_{h}-\frac{\tau_{v}}{t_{cyc}}\frac{\tau_{x}}{t_{cyc}}
           \frac{T_{h}}{4}\left(\frac{d}{ds}\ln \Lambda\right)^2}.
\end{equation}
To consider the entropy production in the relaxation, we rewrite $\dot{\Sigma}$ in Eq.~(\ref{rewrite entropy production rate 1-2}) by using the heat flux in Eq.~(\ref{heat flux expressed by variables}) and the time derivative of the entropy in Eq.~(\ref{definition of entropy}) as follows: 
\begin{equation}
  \label{another relation of the entropy production rate}
  \dot{\Sigma}(t)=\dot{S}(t)-\frac{\dot{Q}(t)}{T(t)}.
\end{equation}
Because the temperature of the heat bath is constant, we derive the entropy production in the relaxation in the hot isothermal process as
\begin{equation}
\label{entropy production in the hot isothermal process in the small relaxation-time regime}
     \int^{\tau_0}_{0}dt\ \dot{\Sigma}=S(\tau_0)-S(0)-\frac{Q^{rel}_h}{T_h},
\end{equation}
where $Q^{rel}_{h}$ is the heat flowing in this relaxation.
In the small relaxation-times regime, the relaxation is very fast (see the Appendix), and the stiffness is regarded to be unchanged in the relaxation because of Eq.~(\ref{stiffness in the relaxation}).
From Eq.~(\ref{sigma_x with small relaxation times}), $\sigma_x$ is also unchanged during the relaxation under the condition of Eq.~(\ref{condition of the adiabatic process to achieve the Carnot efficiency}).
Thus, the heat related to the potential change in Eq.~(\ref{heat derived from potential energy}) in the relaxation vanishes.
By using Eqs.~(\ref{heat expressed by variables}) and (\ref{kinetic energy change in small relaxation times limit}), $Q^{rel}_h$ is evaluated as
\begin{equation}
\label{heat rel h}
    Q^{rel}_{h}\simeq \frac{\Delta T}{2}.
    %\quad
    %Q^{rel}_c\simeq -\frac{\Delta T}{2},
\end{equation}
In addition because $d(\ln\Lambda)/ds$ is noninfinite, as shown in the Appendix, we can approximate the entropy in Eq.~(\ref{definition of entropy}) after the relaxation by
\begin{equation}
  \label{entropy approximation}
  S(t) \simeq \frac{1}{2}\ln(T^2(t))
  + \frac{1}{2}\ln\left(\frac{4\pi^2}{m\lambda(t)}\right) + 1,
\end{equation}
where we used the approximation
\begin{equation}
  m\lambda(\sigma_{x}\sigma_{v}-\sigma_{xv}^2)
  \simeq  T^2 - \frac{\tau_{x}}{t_{cyc}}\frac{\tau_{v}}{t_{cyc}}
  \frac{T^2}{4}
  \left(\frac{d}{ds}\ln\Lambda\right)^2\simeq T^2,
\end{equation}
from Eqs.~(\ref{assumption of the protocol}) and (\ref{approximation of variables}).
The initial state of the hot isothermal process is given by the final state of the cold isothermal process as
\begin{equation}
\label{state D}
\sigma_{x} \simeq \frac{T_{c}}{\lambda_{D}},\quad\sigma_{v} \simeq \frac{T_{c}}{m},\quad\sigma_{xv} \simeq -\frac{T_{c}}{2\lambda_{D}^2}\left.\frac{d\lambda}{dt}\right|_{t=t_{cyc}-0},
\end{equation}
from Eq.~(\ref{approximation of variables}).
Because the stiffness remains $\lambda_{A}$ in the relaxation,  the variables relax to the following values: 
\begin{equation}
\label{state A}
 \sigma_{x} \simeq \frac{T_{h}}{\lambda_{A}},  \quad  \sigma_{v} \simeq \frac{T_{h}}{m},  \quad  \sigma_{xv} \simeq -\frac{T_{h}}{2\lambda_{A}^2}\left.\frac{d\lambda}{dt}\right|_{t=0+0},
\end{equation}
from Eq.~(\ref{approximation of variables}).
%Before the relaxation the temperature and stiffness are $T_c$ and $\lambda_D$ respectively, and they satisfy $T_h$ and $\lambda_A$ after the relaxation.
Using Eqs.~(\ref{entropy approximation})--(\ref{state A}), the difference between %Eq.~(\ref{definition of entropy}), 
$S(0)$ and $S(\tau_{0})$ can be approximated by
\begin{equation}
\label{entropy difference approximation}
    S(\tau_0)-S(0)\simeq  \frac{1}{2}\ln(T_{h}^2)-\frac{1}{2}\ln(T_{c}^2)
    +\frac{1}{2}\ln\left(\frac{\lambda_{D}}{\lambda_{A}}\right).
\end{equation}
We can then evaluate the entropy production in the relaxation in Eq.~(\ref{entropy production in the hot isothermal process in the small relaxation-time regime}) as
\begin{align}
  \label{entropy production in the relaxation in the hot isothermal process}
    \int^{\tau_{0}}_{0} dt\ \dot{\Sigma} 
    \simeq & \frac{1}{2}\ln(T_{h}^2)-\frac{1}{2}\ln(T_{c}^2)
    +\frac{1}{2}\ln\left(\frac{\lambda_{D}}{\lambda_{A}}\right)-\frac{\Delta T}{2T_{h}}\nonumber\\
    =&\frac{1}{2}\ln\left(\frac{T_{h}}{T_{c}}\right)-\frac{\Delta T}{2T_{h}},
\end{align}
using Eqs.~(\ref{condition of the adiabatic process to achieve the Carnot efficiency}),
(\ref{edges of the protocol}),
(\ref{heat rel h}), and (\ref{entropy difference approximation}).
Thus, by using Eqs.~(\ref{entropy production hot after the relaxation}) and (\ref{entropy production in the relaxation in the hot isothermal process}), the entropy production in the hot isothermal process in Eq.~(\ref{entropy production in the hot isothermal process}) is given by
\begin{align}
  \label{entropy production in the hot isothermal process 2}
    \Sigma_{h} 
    \simeq& \frac{1}{2}\ln\left(\frac{T_{h}}{T_{c}}\right)-\frac{\Delta T}{2T_{h}}\\
    &+\frac{1}{T_{h}}\int^{t_{h}/t_{cyc}}_{\tau_{0}/t_{cyc}}ds\
    \frac{\frac{\tau_{v}}{t_{cyc}}\left(\frac{dQ(s)}{ds}\right)^2
      +\frac{\tau_{x}}{t_{cyc}}\frac{T_{h}^2}{4}\left(\frac{d}{ds}\ln \Lambda\right)^2}
         {T_{h}-\frac{\tau_{v}}{t_{cyc}}\frac{\tau_{x}}{t_{cyc}}
           \frac{T_{h}}{4}\left(\frac{d}{ds}\ln \Lambda\right)^2}.\nonumber
\end{align}
Similarly, the entropy production
in the cold isothermal process $\Sigma_{c}$ is given by
\begin{align}
  \label{entropy production in the cold isothermal process}
    \Sigma_{c} \equiv& \int^{t_{cyc}}_{t_{h}}dt\ \dot{\Sigma}
    \simeq \frac{1}{2}\ln\left(\frac{T_{c}}{T_{h}}\right)+\frac{\Delta T}{2T_{c}}\\
    +&\frac{1}{T_{c}}\int^{1}_{(t_{h}+\tau_{1})/t_{cyc}}ds\
    \frac{\frac{\tau_{v}}{t_{cyc}}\left(\frac{dQ(s)}{ds}\right)^2
      +\frac{\tau_{x}}{t_{cyc}}\frac{T_{c}^2}{4}\left(\frac{d}{ds}\ln \Lambda\right)^2}
         {T_{c}-\frac{\tau_{v}}{t_{cyc}}\frac{\tau_{x}}{t_{cyc}}
           \frac{T_{c}}{4}\left(\frac{d}{ds}\ln \Lambda\right)^2},\nonumber
\end{align}
where $\tau_{1}\equiv \max(\tau_x(t_{h}+0), \tau_{v})$ is the relaxation time
at the beginning of the cold isothermal process.
Because no entropy production is present in the adiabatic processes, the entropy production $\Sigma$ per cycle
in the small relaxation-times regime is given by
\begin{align}
  \label{entropy production per cycle}
    \Sigma =& \Sigma_{h}+\Sigma_{c}\nonumber\\
    \simeq&\frac{1}{T_{h}}\int^{t_{h}/t_{cyc}}_{\tau_{0}/t_{cyc}}ds\
    \frac{\frac{\tau_{v}}{t_{cyc}}\left(\frac{dQ(s)}{ds}\right)^2
      +\frac{\tau_{x}}{t_{cyc}}\frac{T_{h}^2}{4}\left(\frac{d}{ds}\ln \Lambda\right)^2}
         {T_{h}-\frac{\tau_{v}}{t_{cyc}}\frac{\tau_{x}}{t_{cyc}}
           \frac{T_{h}}{4}\left(\frac{d}{ds}\ln \Lambda\right)^2}\nonumber\\
         +&\frac{1}{T_{c}}\int^{1}_{(t_{h}+\tau_{1})/t_{cyc}}ds\
         \frac{\frac{\tau_{v}}{t_{cyc}}\left(\frac{dQ(s)}{ds}\right)^2
           +\frac{\tau_{x}}{t_{cyc}}\frac{T_{c}^2}{4}\left(\frac{d}{ds}\ln \Lambda\right)^2}
              {T_{c}-\frac{\tau_{v}}{t_{cyc}}\frac{\tau_{x}}{t_{cyc}}
                \frac{T_{c}}{4}\left(\frac{d}{ds}\ln \Lambda\right)^2}\nonumber\\
              +&\frac{(\Delta T)^2}{2T_{h}T_{c}},
\end{align}
using Eqs.~(\ref{entropy production in the hot isothermal process 2}) and (\ref{entropy production in the cold isothermal process}).
\begin{figure}[t]
  \includegraphics[width=6.2cm]{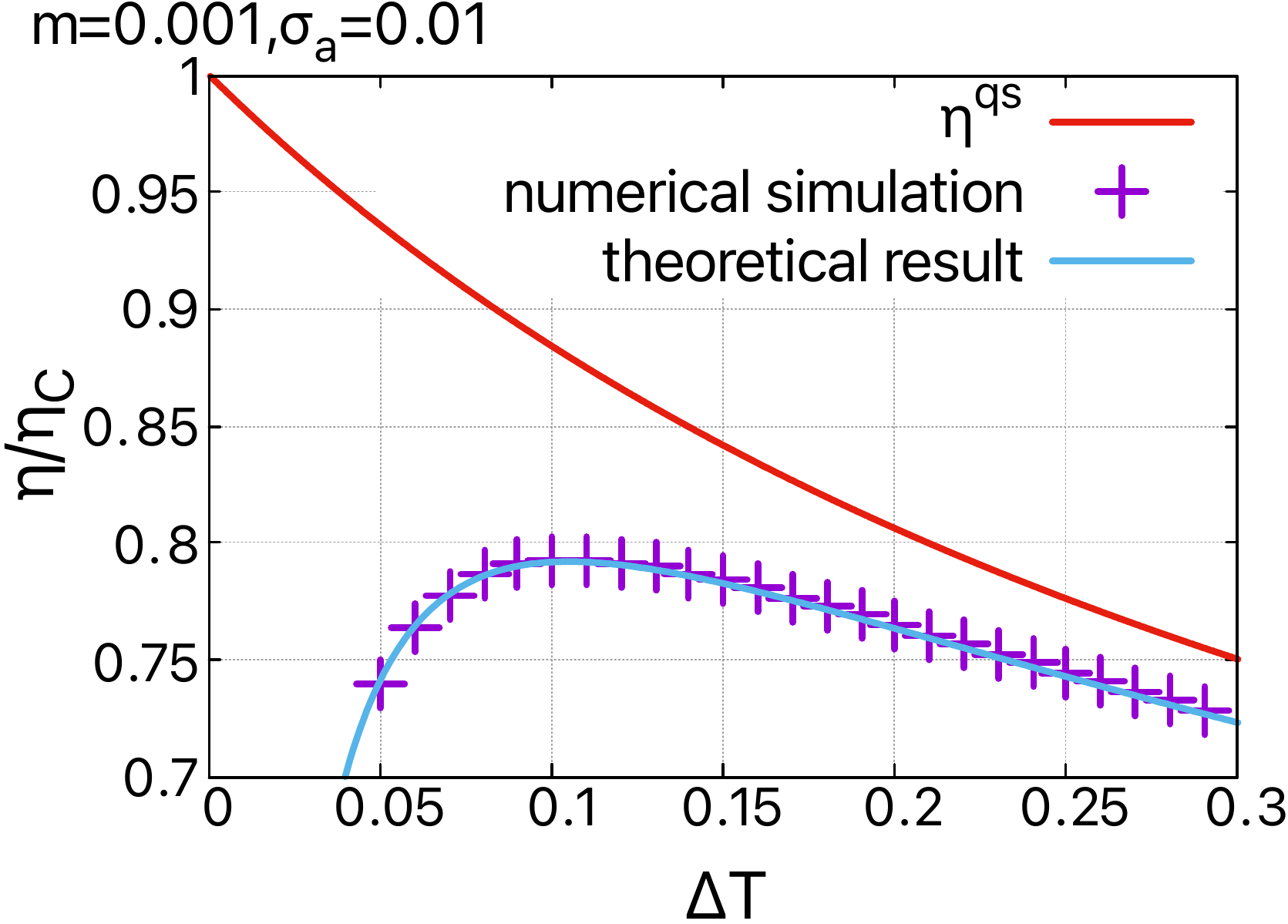}
  \caption{
      The ratio of the efficiency to the Carnot efficiency derived from the numerical simulations in Fig.~\ref{fig:efficiency} in Sec.~\ref{numerical simulation} (purple plus) and theoretical analysis 
      %in Sec.~\ref{theoretical analysis} 
      (sky-blue solid line).
      We set $m=10^{-3}$ and $\sigma_a=10^{-2}$. Although the relaxation times corresponding to these parameters are not very small among the parameters used in Fig.~\ref{fig:efficiency}, the theoretical result and numerical simulations show a good agreement. We have confirmed a better agreement with smaller parameters (data not shown).}
  \label{fig:compare}
\end{figure}

Comparing Eqs.~(\ref{the entropy production with the irreversible works}) and (\ref{entropy production per cycle}),
we can derive the expression of the irreversible works as
\begin{align}
  \label{concrete expression of the irreversible work in the hot isothermal process}
  W^{irr}_{h}
  =& \int^{t_{h}/t_{cyc}}_{\tau_{0}/t_{cyc}}ds\
  \frac{\frac{\tau_{v}}{t_{cyc}}\left(\frac{dQ(s)}{ds}\right)^2
    +\frac{\tau_{x}}{t_{cyc}}\frac{T_{h}^2}{4}\left(\frac{d}{ds}\ln \Lambda\right)^2}
       {T_{h}-\frac{\tau_{v}}{t_{cyc}}\frac{\tau_{x}}{t_{cyc}}
         \frac{T_{h}}{4}\left(\frac{d}{ds}\ln \Lambda\right)^2},\\
  \label{concrete expression of the irreversible work in the cold isothermal process}
  W^{irr}_{c}
  =&\int^{1}_{(t_{h}+\tau_{1})/t_{cyc}}ds\
  \frac{\frac{\tau_{v}}{t_{cyc}}\left(\frac{dQ(s)}{ds}\right)^2
    +\frac{\tau_{x}}{t_{cyc}}\frac{T_{c}^2}{4}\left(\frac{d}{ds}\ln\Lambda\right)^2}
       {T_{c}-\frac{\tau_{v}}{t_{cyc}}\frac{\tau_{x}}{t_{cyc}}
         \frac{T_{c}}{4}\left(\frac{d}{ds}\ln \Lambda\right)^2}.
\end{align}
As shown in the Appendix, $dQ/ds$ and $d(\ln \Lambda)/ds$ are noninfinite after the relaxation.
Thus, the entropy production rate in Eq.~(\ref{rewrite entropy production rate 2}) after the relaxation vanishes in the vanishing limit of the relaxation times.
From Eqs.~(\ref{concrete expression of the irreversible work in the hot isothermal process}) and (\ref{concrete expression of the irreversible work in the cold isothermal process}), it turns out that the integrand of $W^{irr}_{h,c}$, which is $T_{h,c}\dot{\Sigma}$, vanishes at any $s$ in the vanishing limit of the relaxation times, and the irreversible works also vanish.
Therefore, we can confirm that the efficiency in Eq.~(\ref{efficiency 2}) approaches the quasistatic efficiency in Eq.~(\ref{quasistatic efficiency under the adiabatic condition}) in this limit, theoretically explaining the results of the numerical simulations.
Figure \ref{fig:compare} compares the efficiency obtained from the numerical simulations in Fig.~\ref{fig:efficiency} and the efficiency derived from the theoretical analysis in the small relaxation-times regime.
Here, the efficiency of the theoretical analysis was derived by calculating the irreversible works in Eqs.~(\ref{concrete expression of the irreversible work in the hot isothermal process}) and (\ref{concrete expression of the irreversible work in the cold isothermal process}) and substituting them into Eq.~(\ref{efficiency 2}).
Note that we used Eq.~(\ref{heat flux with small relaxation times}) to calculate $dQ/ds$ in  Eqs.~(\ref{concrete expression of the irreversible work in the hot isothermal process}) and (\ref{concrete expression of the irreversible work in the cold isothermal process}).
We can see that the theoretical result and numerical simulations show a good agreement.

We provide a qualitative explanation for the behavior of the efficiency in Figs.~\ref{fig:efficiency} and \ref{fig:compare}, as below.
We consider the case that the relaxation times are small but finite.
Then, from the above discussion, $W^{irr}_{h}$ and $W^{irr}_{c}$ are positive and small.
When $\Delta T$ is large, $\Delta T\Delta S^{qs}$ in the numerator of Eq.~(\ref{efficiency 2}) is sufficiently larger than $W^{irr}_{h,c}$ since we use the protocol satisfying $\Delta S^{qs}=O(1)$ in the numerical simulation.
Since $T_h$ is larger than $\Delta T$, $T_h\Delta S^{qs}$ in the denominator of Eq.~(\ref{efficiency 2}) is also sufficiently larger than $W^{irr}_{h,c}$.
Thus, the efficiency should mainly depend on $T_h$, $\Delta T$, and $\Delta S^{qs}$ as shown in Eq.~(\ref{efficiency 2}).
Although the efficiency is smaller than the Carnot efficiency because of $\Delta T/2$ due to the heat leakage in the denominator of Eq.~(\ref{efficiency 2}), the heat leakage becomes small and the efficiency increases toward the Carnot efficiency as $\Delta T$ becomes small.
At the same time, however, the irreversible works can be comparable to $\Delta T \Delta S^{qs}$.
From Eq.~(\ref{optimized protocol in the overdamped regime}), the stiffness in each isothermal process depends only on the corresponding temperature.
Since $dQ/ds$ in Eqs.~(\ref{concrete expression of the irreversible work in the hot isothermal process}) and (\ref{concrete expression of the irreversible work in the cold isothermal process}) is evaluated by the protocol as shown in Eq.~(\ref{heat flux with small relaxation times}), $W^{irr}_{h,c}$ depend only on the temperature of each isothermal process, but do not depend on $\Delta T$ in the lowest order of $\Delta T$.
Thus, the irreversible works maintain finite even when $\Delta T$ vanishes.
Then, $\Delta T \Delta S^{qs}$ in Eq.(\ref{efficiency 2}) approaches zero while $W^{irr}_{h,c}$ are positively finite.
Thus, the efficiency turns from increase to decrease as $\Delta T$ becomes small and takes the maximum for a specific value of $\Delta T$ as shown in Figs.~\ref{fig:efficiency} and \ref{fig:compare}.

By using $s=t/t_{cyc}$,
the quantity $\phi(t)$ in Eq.~(\ref{time dependence of temperature phi})
can be expressed as
\begin{equation}
  \phi(s) = \left\{
  \begin{split}
    &1 \hspace{3mm}(0< s < t_{h}/t_{cyc})\\
    &0 \hspace{3mm}(t_{h}/t_{cyc}< s < 1).
  \end{split}
  \right.
\end{equation}
Thus, $\chi$ in Eq.~(\ref{amplitude}) is rewritten by using the relaxation time
of the velocity as
\begin{equation}
  \label{rewrite amplitude}
  \chi = \frac{1}{t_{cyc}}\frac{t_{cyc}}{\tau_{v}}\left(\frac{1}{T_{c}}
  \int^{1}_{0}ds\ \frac{m\sigma_{v}\phi^2}{1-\eta_{C}\phi}\right)
  = \frac{C}{t_{cyc}}\frac{t_{cyc}}{\tau_{v}},
\end{equation}
where $C$ is a positive constant given by
\begin{equation}
  C\equiv \frac{1}{T_{c}}\int^{1}_{0}ds\
  \frac{m\sigma_{v}\phi^2}{1-\eta_{C}\phi}.
\end{equation}
In the relaxation at the beginning of each isothermal process, $m\sigma_v$ is positively finite.
After the relaxation, $m\sigma_v$ is approximated by the temperature of the heat bath.
Thus, $C$ is positively finite.
From Eq.~(\ref{rewrite amplitude}),
$\chi$ turns out to diverge in the limit of $\tau_{v}/t_{cyc}\to 0$ when $t_{cyc}$ is finite.
Although $\tau_{v}/t_{cyc}\to 0$ is satisfied even when $t_{cyc}$ diverges and $\tau_v$ is maintained finite, we do not consider that case because it is in the quasistatic limit.
Using  Eqs.~(\ref{entropy production per cycle})
and (\ref{rewrite amplitude}), we can obtain
$\chi\Sigma$ as follows: 
\begin{align}
  \label{finiteness of the upper bound}
    \chi\Sigma
    \simeq&
    \frac{C}{t_{cyc}T_{h}}\int^{t_{h}/t_{cyc}}_{\tau_0/t_{cyc}}ds\ 
    \frac{\left(\frac{dQ}{ds}\right)^2+
      \frac{\tau_{x}T_{h}^2}{4\tau_{v}}
      \left(\frac{d}{ds}\ln \Lambda\right)^2}{T_{h}
      -\frac{\tau_{v}}{t_{cyc}}\frac{\tau_{x}}{t_{cyc}}\frac{T_{h}}{4}
      \left(\frac{d}{ds}\ln \Lambda\right)^2}\nonumber\\
    &+\frac{C}{t_{cyc}T_{c}}\int^{1}_{(t_{h}+\tau_1)/t_{cyc}}ds\ 
    \frac{\left(\frac{dQ}{ds}\right)^2+
      \frac{\tau_{x}T_{c}^2}{4\tau_{v}}
      \left(\frac{d}{ds}\ln \Lambda\right)^2}{T_{c}
      -\frac{\tau_{v}}{t_{cyc}}\frac{\tau_{x}}{t_{cyc}}\frac{T_{c}}{4}
      \left(\frac{d}{ds}\ln \Lambda\right)^2}\nonumber\\
    &+\frac{C}{\tau_{v}}\frac{(\Delta T)^2}{2T_{h}T_{c}}.
\end{align}

\begin{figure}[t]
  \includegraphics[width=6.2cm]{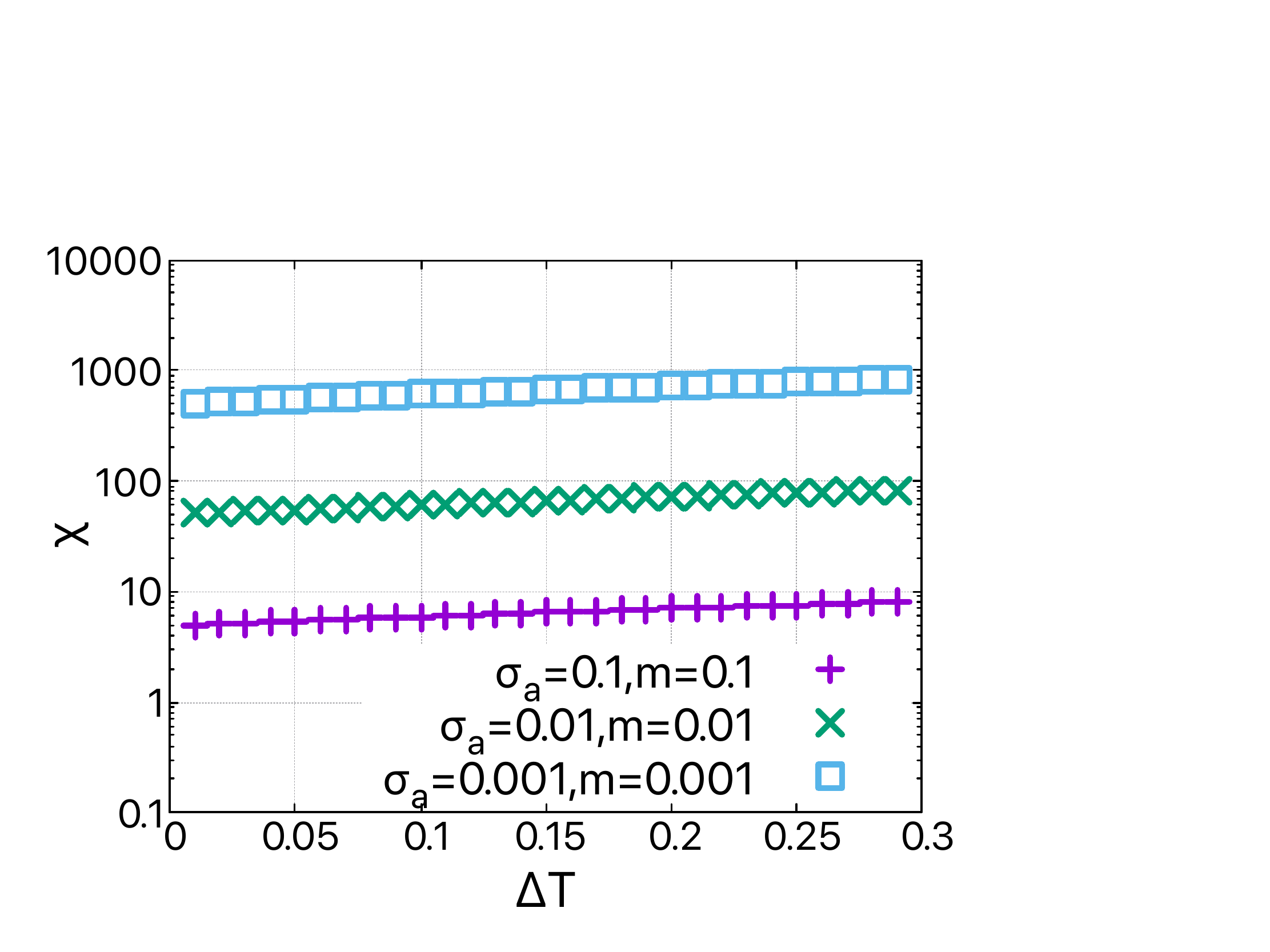}
  \includegraphics[width=6.2cm]{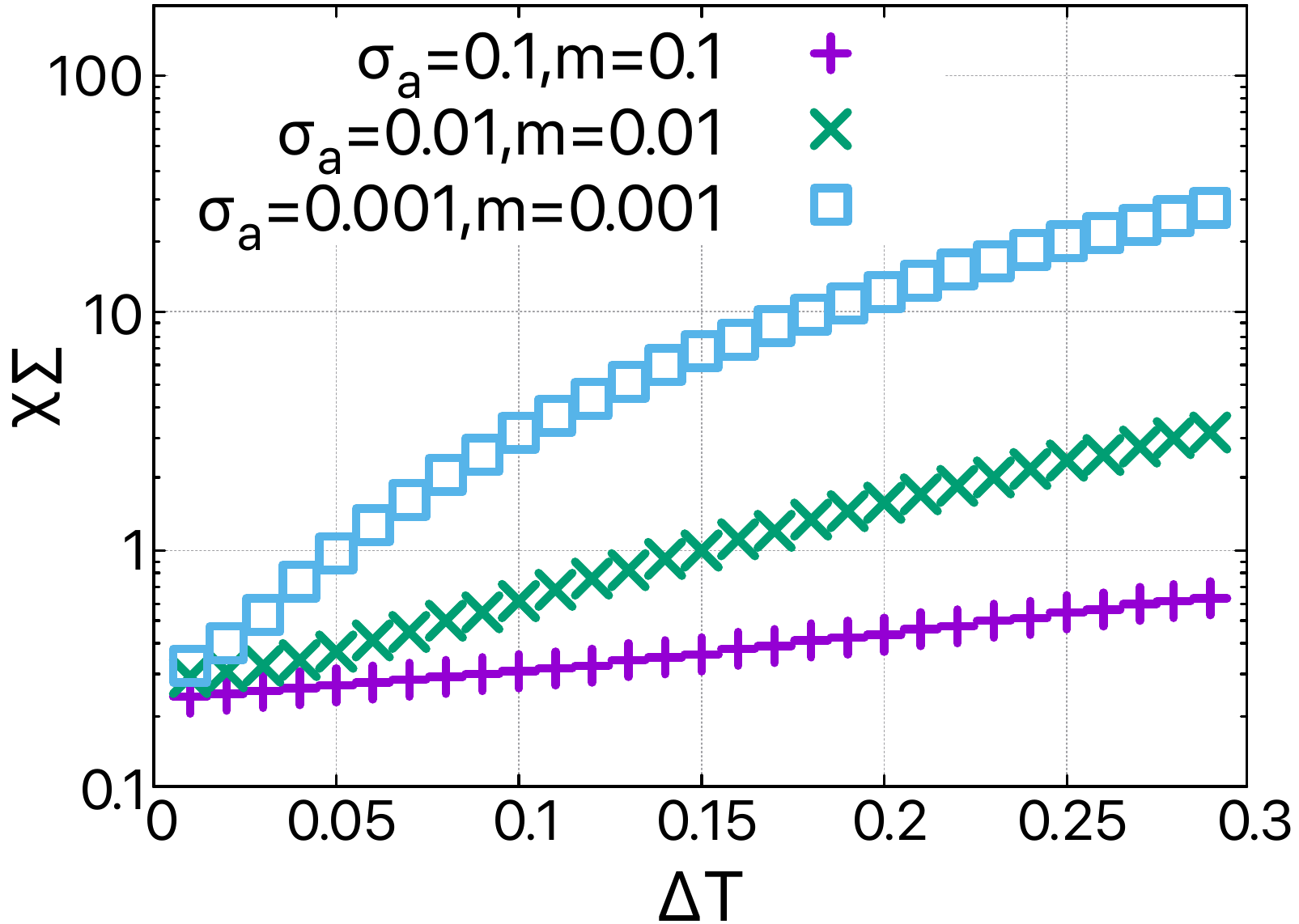}
  \caption{The quantities $\chi$ in Eq.~(\ref{amplitude}) and $\chi\Sigma$
    when $\tau_{x}$ and $\tau_{v}$ are varied.
    Because the parameter $\sigma_{a}$ is proportional to $\tau_{x}(0)$ in the protocol in Eq.~(\ref{optimized protocol in the overdamped regime}), we vary $\sigma_{a}$ to make $\tau_x$ be small.
    Similarly, we vary the mass $m$ because it is proportional to $\tau_{v}$.
    In these simulations, we used $(\sigma_{a}=0.1,m=0.1)$ (purple plus),
    $(\sigma_{a}=0.01,m=0.01)$ (green cross), and $(\sigma_{a}=0.001,m=0.001)$ (sky-blue square).
    We can see that $\chi$ diverges at each $\Delta T$ when we consider the limit of
    $\sigma_{a},m\to 0$ ($\tau_{x},\tau_{v} \to 0$).
    In addition, we can also see that the values of $\chi \Sigma$
    are positively finite for the vanishing limit of $\Delta T$ for any relaxation times.}
  \label{fig:bound}
\end{figure}

Here, we consider the vanishing limit of the relaxation times in the small temperature-difference regime and evaluate the efficiency and power in this limit.
As seen in Eq.~(\ref{efficiency with entropy production}), the efficiency approaches the Carnot efficiency when $\Sigma$ vanishes. Moreover, we evaluate $\Sigma$ in the vanishing limit of $\tau_{x}$ and $\tau_{v}$ in the small temperature-difference regime.
In this limit, we can show that $dQ/ds$ and
$d(\ln\Lambda)/ds$ in Eq.~(\ref{entropy production per cycle}) do not diverge after the relaxation (see the Appendix).
Thus, when the relaxation times vanish at any instant after the relaxation, the entropy production rate always vanishes from Eq.~(\ref{rewrite entropy production rate 2}), and the first and second terms on the right-hand side of Eq.~(\ref{entropy production per cycle}) also vanish.
In addition, when $\Delta T$ is small, the third term in Eq.~(\ref{entropy production per cycle}), which is due to the relaxation, is $O[(\Delta T)^2]$ and can be ignored.
Therefore, the entropy production per cycle in Eq.~(\ref{entropy production per cycle})
should be $O[(\Delta T)^2]$, and the efficiency can be regarded as the Carnot efficiency because of the reasoning presented below Eq.~(\ref{efficiency with entropy production}).
Then, because $dQ/ds$ and $d(\ln\Lambda)/ds$ are always noninfinite, the first and second terms on the right-hand side of Eq.~(\ref{finiteness of the upper bound}) are positively finite in the vanishing limit of $\tau_x$ and $\tau_v$.
Even when $\Delta T$ is small, $\chi\Sigma$ is positive,
and the right-hand side of the trade-off relation in Eq.~(\ref{trade-off relation in the Brownian Carnot cycle}) is positive.
Therefore, the finite power may be allowed
even when $\Sigma$ vanishes.
In the above limit, because the irreversible works in Eqs.~(\ref{concrete expression of the irreversible work in the hot isothermal process}) and (\ref{concrete expression of the irreversible work in the cold isothermal process}) vanish, the power in Eq.~(\ref{power}) approaches $P^{*}$ in Eq.~(\ref{quasistatic power}), which implies that the power is finite.
Therefore, the Carnot efficiency is achievable in the finite-power Brownian Carnot cycle
without breaking the trade-off relation in Eq.~(\ref{trade-off relation in the Brownian Carnot cycle}).

In Fig.~\ref{fig:bound}, we numerically confirmed that $\chi$ increases and $\chi\Sigma$ remains positively finite in the limit of $\Delta T\to 0$ when we consider smaller relaxation times.
We can expect $\chi$ to diverge while maintaining $\chi\Sigma$ positively finite in the vanishing limit of the relaxation times in the limit of $\Delta T\to 0$.
This result implies that $\Sigma$ vanishes while maintaining $\chi\Sigma$ positively finite, and we can expect that $\Sigma$ vanishes and $\chi$ diverges simultaneously in the vanishing limit of the relaxation times.
%Note that the third term of Eq.~(\ref{finiteness of the upper bound}) is not negligible and diverges in the vanishing limit of $\tau_v$ when $\Delta T$ is large.

\section{Summary and discussion}
\label{summary and discussion}
Motivated by the previous study~\cite{PhysRevLett.121.120601}, we studied the relaxation-times dependence of the efficiency and power in a Brownian Carnot cycle with the instantaneous adiabatic processes and  time-dependent harmonic potential, described by the underdamped Langevin equation.
In this system, we numerically showed that the Carnot efficiency is compatible with finite power in the vanishing limit of the relaxation times in the small temperature-difference regime.
We analytically showed that the present results are consistent with the trade-off relation between efficiency and power, which was proved for more general systems in
\cite{PhysRevLett.117.190601,PhysRevE.96.022138,PhysRevE.97.062101}.
By expressing the trade-off relation using the entropy production in terms of the relaxation times of the system, we demonstrated that such compatibility is possible by both the diverging constant and the vanishing entropy production in the trade-off relation in the vanishing limit of the relaxation times.

In the numerical simulation results in Sec.~\ref{numerical simulation},
we used a specific protocol.
However, we can use other protocols satisfying the following three conditions
to achieve the Carnot efficiency and finite power
in the small temperature-difference regime.
The first condition is that the protocol should satisfy
the condition in
Eq.~(\ref{condition of the adiabatic process to achieve the Carnot efficiency}).
%Eqs.~(\ref{condition of the adiabatic process (ii)})
%and (\ref{condition of adiabatic process (iv)}).
For such a protocol,
the heat leakage in the relaxation at
the beginning of the isothermal processes is $O(\Delta T)$.
Thus, heat leakage can be neglected
in the small temperature-difference regime,
compared with the heat flowing in the isothermal processes.
The second condition is that 
the stiffness is expressed by using
a scaling function as in Eq.~(\ref{assumption of the protocol}).
The third condition of the protocols is that
the stiffness diverges at any instant of time.
This is satisfied by the vanishing relaxation time of position,
and it is one of the necessary conditions for the entropy production rate vanishing after the relaxation,
%in the vanishing limit of the relaxation times
as we showed in Sec.~\ref{theoretical analysis}.
When the entropy production rate at any instant vanishes,
irreversible works also vanish, which allows us to derive the compatibility of the Carnot efficiency and finite power
in the small temperature-difference regime.

Note that we showed that achieving both the Carnot efficiency and finite power is possible in the small temperature-difference regime without breaking the trade-off relation in Eq.~(\ref{trade-off relation in our cycle}) of the proposed cycle.
In the linear irreversible thermodynamics,
which can describe the heat engines operating in the small temperature-difference regime,
the currents of the systems are described by the linear combination of affinities,
and their coefficients are called the Onsager coefficients.
When these coefficients have the reciprocity resulting from
the time-reversal symmetry of the systems,
a previous study~\cite{PhysRevLett.106.230602} showed
that the compatibility of the Carnot efficiency with finite power is forbidden.
The same study also showed that the compatibility can be allowed
in the systems without time-reversal symmetry.
However, in some studies related to the concrete systems without time-reversal
symmetry~\cite{PhysRevLett.110.070603,PhysRevB.87.165419,
PhysRevB.94.121402,PhysRevLett.112.140601,
PhysRevLett.114.146801,Sothmann_2014}, the compatibility has not been found thus far.
On the other hand, there is a possibility of the compatibility
of the Carnot efficiency and finite power when
the Onsager coefficients with reciprocity show diverging behaviors (cf. Eq. (7) in Ref.~\cite{N.Shiraishi2018}).
The Onsager coefficients of our Carnot cycle can be obtained in the same way as Ref.~\cite{Izumida2010}, which have reciprocity.
In the vanishing limit of the relaxation times, we can show the divergence of these Onsager coefficients.
Although the effect of the asymmetric limit of the non-diagonal Onsager coefficients on 
the linear irreversible heat engines realizing the Carnot efficiency at finite power was studied in Ref.~\cite{PhysRevLett.106.230602},
this case is different from our case where all of the Onsager coefficients show the diverging behaviors.

Furthermore, another study reported the compatibility of the Carnot efficiency with finite power using a time-delayed system within the linear response theory~\cite{Bonan_a_2019}.
Because the time-delayed systems are not described by the Markovian dynamics, the trade-off relation in Eq.~(\ref{general trade-off relation}) may not be applied to them.
Thus, there may be a possibility to achieve the Carnot efficiency in finite-power non-Markovian heat engines.
In this paper, however, we showed that achieving both the Carnot efficiency and finite power is possible in a Markovian heat engine.

Although we have used the instantaneous adiabatic process, the other type of adiabatic process can be used for the study of the Brownian Carnot cycle~\cite{PhysRevE.101.032129,Plata_2020,PhysRevLett.114.120601,PhysRevLett.121.120601}. 
In this adiabatic process, the system contacts with a heat bath with varying temperature that maintains vanishing heat flow between the system and the heat bath on average. 
While the Brownian Carnot cycle utilizing this adiabatic process does not suffer from the heat leakage, mathematical treatment may become more difficult. 
Therefore, it is a challenging task to study the detailed relaxation-times dependence of the efficiency and power for this cycle, which we will report elsewhere.

\section*{acknowledgments}
We thank S.-i. Sasa and Y. Suda for their helpful discussions.

\appendix
\def\thesection{}
\section{Behavior of heat flux in the vanishing limit of relaxation times}
\renewcommand{\theequation}{A\arabic{equation}}

\label{Behavior of heat flux in the vanishing limit of relaxation times}
We show that heat flux $\dot{Q}$ after the relaxation in an isothermal process is noninfinite in the vanishing limit of the relaxation times.
For this purpose, we first consider the case where
the stiffness $\lambda$ and the temperature $T$ are constant.
We assume that an isothermal process lasts for $t_{i} < t < t_{f}$.
%As discussed in Sec.~\ref{Carnot cycle},
%we assumed that the stiffness can be expressed as $\lambda(t)=\Lambda(s)$
%by using $s\equiv t/t_{cyc}$, where $t_{cyc}$ is the cycle time.
As the adiabatic processes take no time,
the variables $\sigma_{x}$, $\sigma_{v}$, and $\sigma_{xv}$
at the beginning of the isothermal process
should be unchanged from the end of the preceding isothermal process.
We set $\sigma_{x}(t_{i})=\sigma_{x0}$, $\sigma_{v}(t_{i})=\sigma_{v0}$, and $\sigma_{xv}(t_{i})=\sigma_{xv0}$.
Under these initial conditions, we can solve 
Eqs.~(\ref{equation of sigma_x})--(\ref{equation of sigma_xv})
using the Laplace transform~\cite{PhysRevE.97.022131},
and we can obtain $\sigma_{x}$ and $\sigma_{v}$ as follows:
\begin{equation}
  \label{sigma_x with constant lambda}
  \begin{split}
    \sigma_{x}(t) =&\frac{T}{\lambda}
    +\frac{m}{\lambda} D_{1}e^{-\frac{\gamma}{m}(t-t_{i})}\\
    &+\frac{\left(\gamma + m\omega^{*}\right)^2}{4\lambda^2}
    D_{2}e^{-\left(\frac{\gamma}{m} - \omega^{*}\right)(t-t_{i})}\\
    & +\frac{\left(\gamma - m\omega^{*}\right)^2}{4\lambda^2}
    D_{3}e^{-\left(\frac{\gamma}{m} + \omega^{*}\right)(t-t_{i})},
  \end{split}
\end{equation}
\begin{equation}
  \label{sigma_v with constant lambda}
  \begin{split}
    \sigma_{v}(t)=&\frac{T}{m}
    +D_{1}e^{-\frac{\gamma}{m} (t-t_{i})}
    +D_{2}e^{-\left(\frac{\gamma}{m} - \omega^{*}\right)(t-t_{i})}\\
    &+D_{3}e^{-\left(\frac{\gamma}{m} + \omega^{*}\right)(t-t_{i})},
  \end{split}
\end{equation}
where
\begin{align}
  \label{omega star}
  \omega^{*} \equiv& \frac{\gamma}{m}\sqrt{1-4\frac{m\lambda}{\gamma^2}},
\end{align}
\begin{align}
  \label{D1}
  D_{1} \equiv &\frac{\lambda}{m{\omega^{*}}^2}
  \left(4\frac{T}{m}-2\sigma_{v0}
  -2\frac{\lambda}{m}\sigma_{x0}
  -2\frac{\gamma}{m}\sigma_{xv0}
  \right),
\end{align}
%\end{equation}
\begin{align}
  \label{D2}
  %\begin{split}
    D_{2} \equiv &-\frac{1}{2{\omega^{*}}^2}\left[\frac{\gamma T}{m^2}\left(\frac{\gamma}{m}-\omega^{*}\right)
      + \left( 2\frac{\lambda}{m}-\frac{\gamma^2}{m^2}+\frac{\gamma}{m}\omega^{*}\right)\sigma_{v0}\right.\nonumber\\
      &-2\frac{\lambda^2}{m^2}\sigma_{x0}
      \left.+2\frac{\lambda}{m}\left(-\frac{\gamma}{m}+\omega^{*}\right)\sigma_{xv0}\right],
  %\end{split}
\end{align}
\begin{align}
  \label{D3}
  %\begin{split}
    D_{3} \equiv &-\frac{1}{2{\omega^{*}}^2}\left[\frac{\gamma T}{m^2}
      \left(\frac{\gamma}{m}+\omega^{*}\right)
      + \left(2\frac{\lambda}{m}-\frac{\gamma^2}{m^2}-\frac{\gamma}{m}\omega^{*}\right)\sigma_{v0}\right.\nonumber\\
      &-2\frac{\lambda^2}{m^2}\sigma_{x0}
      \left.+2\frac{\lambda}{m}\left(-\frac{\gamma}{m} - \omega^{*}\right)\sigma_{xv0}\right].
  %\end{split}
\end{align}
We can also derive $\sigma_{xv}$
using Eqs.~(\ref{equation of sigma_xv}) and (\ref{sigma_x with constant lambda}).
We can rewrite Eq.~(\ref{omega star}) using the relaxation times
Eqs.~(\ref{relaxation time of x}) and (\ref{relaxation time of v}) as
\begin{equation}
  \label{omega star with relaxation times}
  \omega^{*} = \frac{1}{\tau_{v}}\sqrt{1-4\frac{\tau_{v}}{\tau_{x}}}.
\end{equation}
The exponential functions in Eqs.~(\ref{sigma_x with constant lambda})
and (\ref{sigma_v with constant lambda}) are represented using $s= t/t_{cyc}$ and the relaxation times as
\begin{align}
  \label{exponential term 1}
  e^{-\frac{\gamma}{m} (t-t_{i})}
  &= e^{-\frac{t_{cyc}}{\tau_{v}}\left(s-\frac{t_{i}}{t_{cyc}}\right)},\\
  \label{exponential term 2}
  e^{-\left(\frac{\gamma}{m} - \omega^{*}\right)(t-t_{i})}
  &= e^{-\frac{t_{cyc}}{\tau_{v}}\left(1-\sqrt{1-4\frac{\tau_{v}}{\tau_{x}}}\right)
    \left(s-\frac{t_{i}}{t_{cyc}}\right)},\\
  \label{exponential term 3}
  e^{-\left(\frac{\gamma}{m} + \omega^{*}\right)(t-t_{i})}
  &= e^{-\frac{t_{cyc}}{\tau_{v}}\left(1+\sqrt{1-4\frac{\tau_{v}}{\tau_{x}}}\right)
    \left(s-\frac{t_{i}}{t_{cyc}}\right)}.
\end{align}
When we consider $\tau_{x} \le 4\tau_{v}$,
$\omega^{*}$ in Eq.~(\ref{omega star with relaxation times}) becomes purely imaginary.
Thus, the exponential terms in Eqs.~(\ref{exponential term 1})--(\ref{exponential term 3})
vanish in $\tau_{v}/t_{cyc}\to 0$ when $s>t_{i}/t_{cyc}$ is satisfied.
When we also consider $\tau_{x} > 4\tau_{v}$,
the exponential terms in Eqs.~(\ref{exponential term 1}) and (\ref{exponential term 3}), 
vanish in $\tau_{v}/t_{cyc}\to 0$.
Because the exponent of Eq.~(\ref{exponential term 2}) can be approximated by
\begin{equation}
  -\frac{t_{cyc}}{\tau_{v}}\left(1-\sqrt{1-4\frac{\tau_{v}}{\tau_{x}}}\right)
    \left(s-\frac{t_{i}}{t_{cyc}}\right)
  \simeq -2\frac{t_{cyc}}{\tau_{x}}\left(s-\frac{t_{i}}{t_{cyc}}\right),
\end{equation}
the exponential terms in Eq.~(\ref{exponential term 2}) vanishes in $\tau_{x}/t_{cyc}\to 0$ when $s>t_{i}/t_{cyc}$ is satisfied.
Thus, the exponential terms vanish
in any value of $\tau_{v}/\tau_{x}$
in the vanishing limit of $\tau_x$ and $\tau_v$
when $s-t_{i}/t_{cyc}$ is positively finite.
Therefore, $\sigma_{x}$ and $\sigma_{v}$ after the relaxation are approximated by 
\begin{equation}
  \label{sigma_x with small relaxation times}
  \sigma_{x} \simeq \frac{T}{\lambda},
    \quad
  \sigma_{v} \simeq \frac{T}{m}.
\end{equation}
To obtain $\sigma_{xv}$, we use Eqs.~(\ref{equation of sigma_x}) and (\ref{sigma_x with constant lambda}) as follows: 
Because $T$ and $\lambda$ are constant, the time derivative of the first term in Eq.~(\ref{sigma_x with constant lambda}) disappears.
Moreover, as the exponential terms vanish rapidly,
the remaining terms in Eq.~(\ref{sigma_x with constant lambda})  vanish after the relaxation even when we differentiate them with respect to time.
Thus $\sigma_{xv}$ vanishes after the relaxation.
\begin{figure}[t]
  \includegraphics[width=6.1cm]{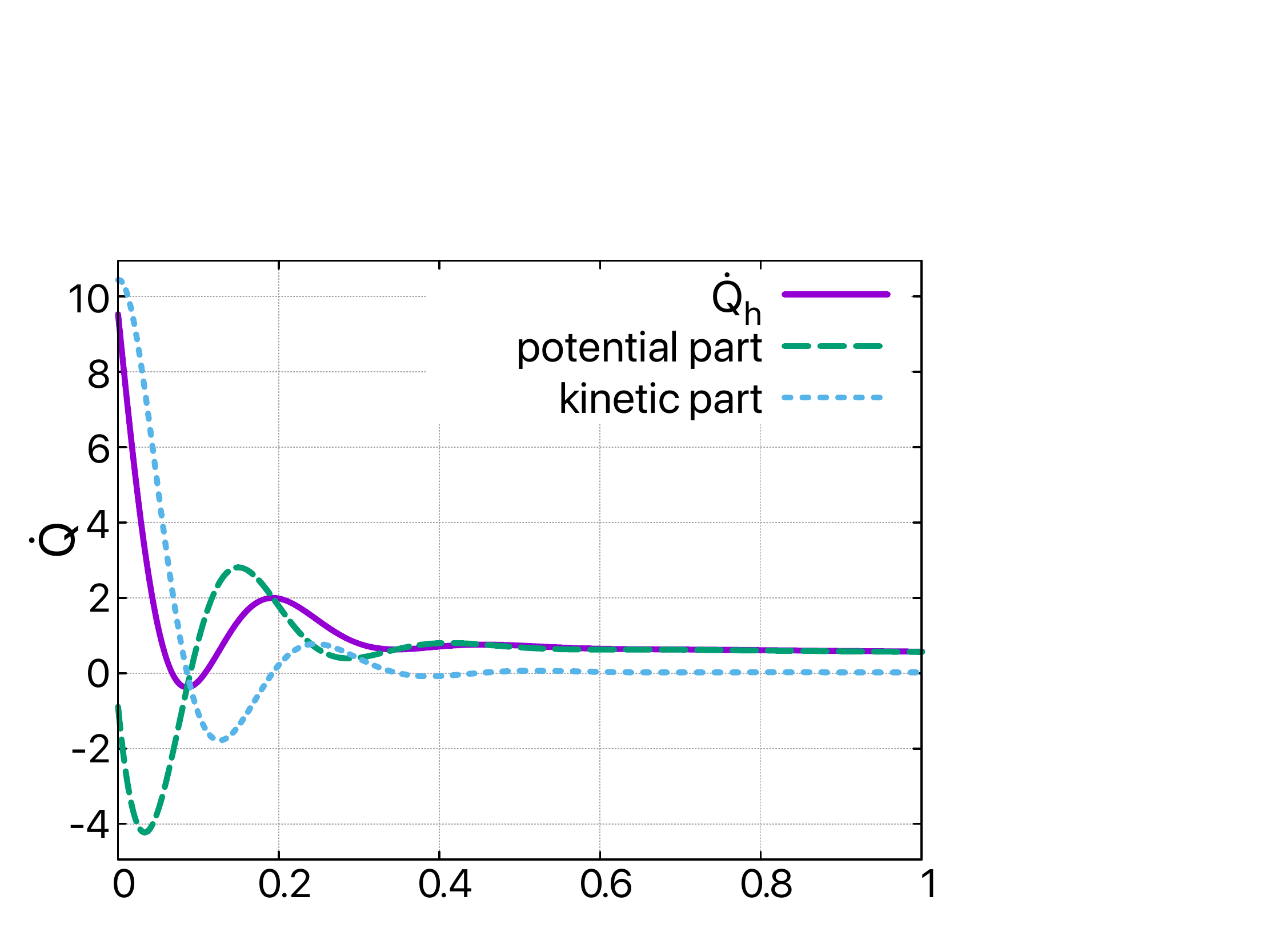}
  \includegraphics[width=6.1cm]{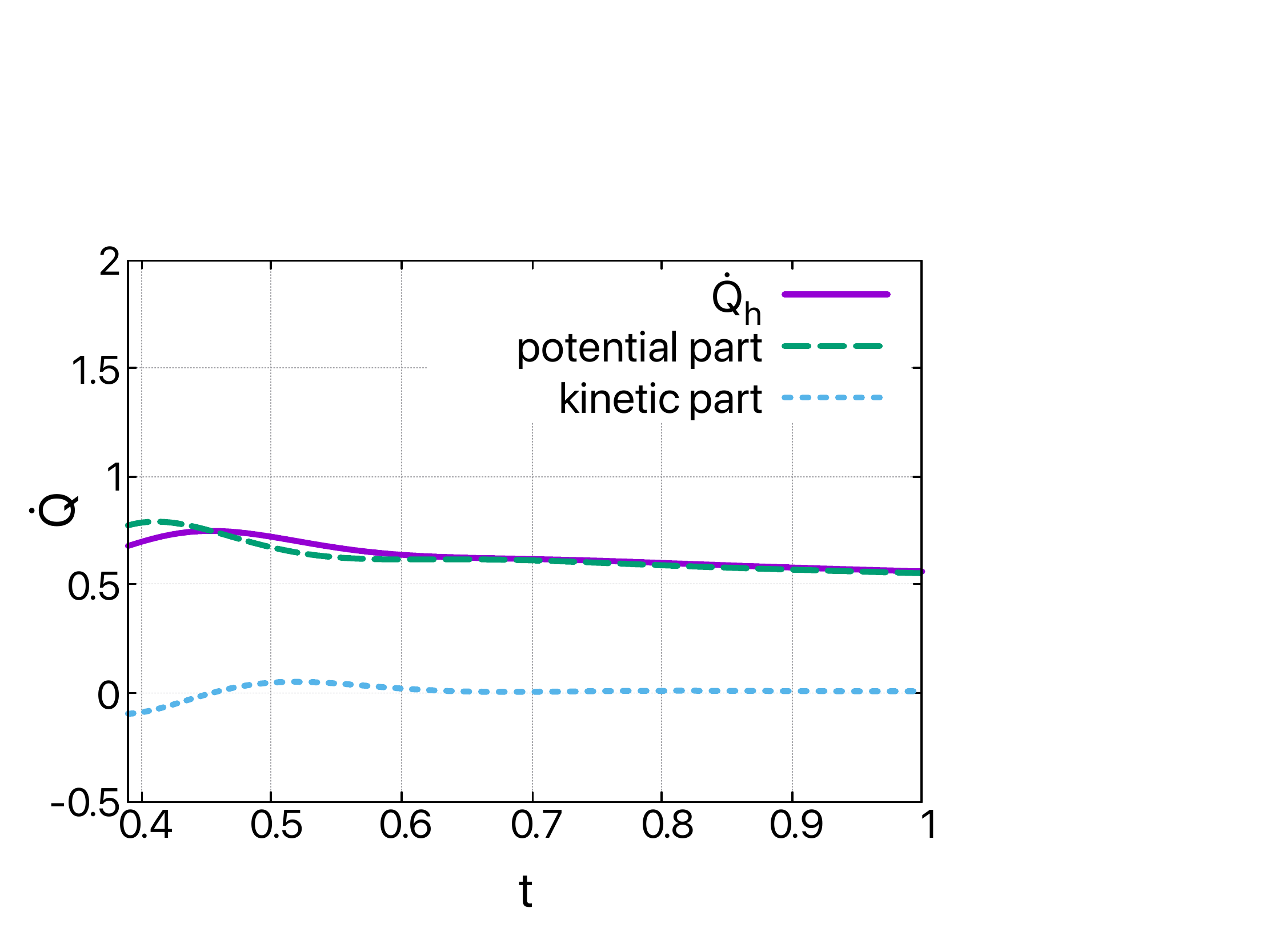}
  \caption{Time evolution of $\dot{Q}_{h}(t)$ (purple solid line),
    its potential part $\lambda \dot{\sigma}_{x}/2$ (green dashed line),
    and its kinetic part $m\dot{\sigma}_{v}/2$ (sky-blue dotted line)
    in the hot isothermal process.
    We can see a relaxation at the beginning of the process.
    The lower figure is an enlargement view of a part of the upper figure,
    which shows that
    $\dot{Q}_{h}(t) \simeq \lambda(t)\dot{\sigma}_{x}(t)/2$ and $m\dot{\sigma}_{v}(t)\simeq 0$
    are satisfied.
    In this simulation, we used $\lambda(t)$ in
    Eq.~(\ref{optimized protocol in the overdamped regime})
    and set $T_{h}=2.0$, $T_{c}=1.0$, $t_{h}=t_{c}=1.0$,
    $m=0.1$, $\sigma_{a}=0.1$, $\gamma =1.0$, and $\sigma_{b}/\sigma_{a}=2.0$.
  }
  \label{fig:efficiency whose relaxation time is const.}
\end{figure}

Subsequently, we consider the isothermal process where the stiffness $\lambda$ depends on time.
When $\tau_j$ ($j=x,v$) is sufficiently small, by using the Taylor expansion, we derive
\begin{equation}
\label{stiffness in the relaxation}
\begin{split}
    \lambda(t+\tau_j)\simeq%& \lambda(t)+\tau_j\frac{d\lambda(t)}{dt}
    \lambda(t)\left(1+\tau_j\frac{d}{dt}\ln\lambda(t)\right).
    %\simeq \lambda(t).
\end{split}
\end{equation}
If $d(\ln\lambda)/dt$ is noninfinite in the vanishing limit of $\tau_j$, we can obtain
\begin{equation}
\label{stiffness in the relaxation}
       \lambda(t+\tau_j) \simeq \lambda(t),
\end{equation}
which implies that the stiffness $\lambda$ is constant during the relaxation.
We show that $d(\ln\lambda)/dt$ is noninfinite as below.
Because $\lambda$ varies smoothly in the isothermal process, $\lambda$ is differentiable, and we obtain
\begin{equation}
\label{Taylor expansion}
    \frac{\lambda(t+\Delta t)}{\lambda(t)} \simeq 1+\Delta t \frac{d}{dt}\ln\lambda(t),
\end{equation}
where $\Delta t$ is finite but sufficiently small.
Because we assumed that $\lambda(t_f)/\lambda(t_i)$ is finite at any time $t_i$ and $t_f$ in the isothermal process, as mentioned below Eq.~(\ref{assumption of the protocol}) in Sec.~\ref{Carnot cycle}, $d(\ln\lambda(t))/dt$ should be noninfinite.
Thus, as Eq.~(\ref{stiffness in the relaxation}) is satisfied,
we can regard $\lambda$ as a constant in the relaxation even if $\lambda$ varies with time and diverges.
Thus, we can apply $\sigma_{x}$ and $\sigma_{v}$ in Eqs.~(\ref{sigma_x with constant lambda}) and (\ref{sigma_v with constant lambda}) under constant $\lambda$ to the case of varying $\lambda$ in the relaxation.
Then, $\sigma_{x}$ and $\sigma_{v}$ immediately relax in the vanishing limit of $\tau_x$ and $\tau_v$ and satisfy Eq.~(\ref{sigma_x with small relaxation times}) immediately after the relaxation.
When the stiffness changes from $\lambda(t)$ to $\lambda(t+\Delta t)$ after the relaxation, $\sigma_{x}$ and $\sigma_{v}$ immediately relax to Eq.~(\ref{sigma_x with small relaxation times}) with $\lambda=\lambda(t+\Delta t)$ in the limit of $\tau_x(t),\tau_v\to 0$. 
%, $\Delta t$ can be made arbitrarily small.
Thus, when $\tau_x(t)$ and $\tau_v$ vanish at any instant, we can regard that Eq.~(\ref{sigma_x with small relaxation times}) is always satisfied in the isothermal process after the relaxation.
When we consider $\sigma_{xv}$, the time derivative of $T/\lambda$ in Eq.~(\ref{sigma_x with constant lambda}) does not vanish because $\lambda$ varies smoothly.
The remaining terms in Eq.~(\ref{sigma_x with constant lambda}) vanish after the relaxation even when we differentiate them with respect to time because the exponential terms vanish rapidly.
Using Eq.~(\ref{sigma_x with small relaxation times}),
we obtain the time evolution of $\sigma_{x}$ and $\sigma_{v}$ after the relaxation in the isothermal process with the temperature $T$ as
\begin{equation}
  \label{time dependence of sigma_x}
  \dot{\sigma}_{x}(t)\simeq -\frac{T}{\lambda(t)}\left(\frac{d}{dt}\ln \lambda\right),
%\end{equation}
%\begin{equation}
  %\label{time dependence of sigma_v}
  \quad
  \dot{\sigma}_{v} \simeq 0.
\end{equation}
Then, from Eq.~(\ref{equation of sigma_x}), we obtain
\begin{equation}
  \label{sigma_xv with small relaxation times}
  \sigma_{xv}(t)\simeq -\frac{T}{2\lambda(t)}\left(\frac{d}{dt}\ln \lambda\right).
\end{equation}
The heat flux $\dot{Q}(t)$ in Eq.~(\ref{heat flux expressed by variables}) is represented as
\begin{equation}
  \label{heat flux with small relaxation times}
  \begin{split}
    \dot{Q}(t) 
    %=& \frac{1}{2}m\dot{\sigma}_{v}(t) + \frac{1}{2}\lambda(t)\dot{\sigma}_{x}(t)\\
    \simeq & \frac{1}{2}\lambda(t)\dot{\sigma}_{x}(t)
    \simeq  -\frac{T}{2}\left(\frac{d}{dt}\ln \lambda\right),
  \end{split}
\end{equation}
where we used Eq.~(\ref{time dependence of sigma_x}), and $\dot{Q}$ is noninfinite because $d(\ln \lambda)/dt$ is noninfinite.
Note that we obtain
\begin{align}
\label{finiteness of Lambda and dQ/ds}
    \frac{d}{ds}\ln\Lambda=&t_{cyc}\frac{d}{dt}\ln\lambda,\nonumber\\
    \frac{dQ}{ds}=&t_{cyc}\frac{dQ}{dt}
    =-\frac{T}{2}\left(\frac{d}{ds}\ln\Lambda\right),
\end{align}
using $s$ and Eqs.~(\ref{assumption of the protocol}) and (\ref{heat flux with small relaxation times}).
Because $d(\ln \lambda)/dt$ is noninfinite, $d(\ln\Lambda)/ds$ and $dQ/ds$ are also noninfinite after the relaxation when $t_{cyc}$ is finite.

%We here consider the protocol in Eq.~(\ref{optimized protocol in the overdamped regime})
%characterized by the parameters $\sigma_{a}$ and $\sigma_{b}$.
%In the numerical simulations in Sec.~\ref{numerical simulation},
%we varied the value of $\sigma_{a}$
%with keeping the ratio of $\sigma_{a}$ to $\sigma_{b}$ finite,
%which we also see here.

Figure \ref{fig:efficiency whose relaxation time is const.}
shows a time evolution of the heat flux $\dot{Q}_{h}$,
its potential part $\lambda\dot{\sigma}_x/2$, and its kinetic part $m\dot{\sigma}_{v}/2$
in the hot isothermal process with the protocol in Eq.~(\ref{optimized protocol in the overdamped regime}).
In this simulation, we used the same parameters as in Sec.~\ref{numerical simulation}.
From the figure, we can see a relaxation at the beginning of the process.
As implied in Eq.~(\ref{heat flux with small relaxation times}), the heat flux $\dot{Q}_{h}$ is
almost equal to its potential part $\lambda\dot{\sigma}_{x}/2$,
and the kinetic part $m\dot{\sigma}_{v}/2$ almost vanishes
after the relaxation.

\bibliography{ref_of_article2019}

\end{document}